\documentclass[12pt]{article}
\usepackage{epsfig}
\newcommand{\dDk}{\frac{d^Dk}{(2\pi)^D}}
\newcommand{\dDl}{\frac{d^Dl}{(2\pi)^D}}
\newcommand{\dDp}{\frac{d^Dp}{(2\pi)^D}}
\newcommand{\be}{\begin{equation}}
\newcommand{\ee}{\end{equation}}
\newcommand{\msbar}{\overline{\rm MS}}

\newcommand{\eps}{\varepsilon}
\newcommand{\Li}{{\rm Li}}
\newcommand{\pfrac}[2]{\left(\frac{#1}{#2}\right)}
\newcommand{\slk}{k\kern-6pt/\kern.5pt}
\newcommand{\slp}{p\kern-5pt/}
\newcommand{\slq}{q\kern-5.5pt/}
\newcommand{\slv}{v\kern-5pt\raise1pt\hbox{$\scriptstyle/$}\kern1pt}
\newcommand{\replabel}[1]{\vbox to0pt{\vss\hbox to0pt{\raise 24pt
  \hbox to\hsize{\hfill\rm #1}\hss}}}
\newcommand{\MeV}{{\rm\,MeV}}
\newcommand{\GeV}{{\rm\,GeV}}

\newcommand{\Tr}{\mathop{\rm Tr}\nolimits}
\newcommand{\tr}{\mathop{\rm tr}\nolimits}

\newcommand{\ice}[1]{\relax}
\newcommand{\bea}{\begin{eqnarray}}
\newcommand{\eea}{\end{eqnarray}}
\begin{document}
\begin{flushright}
MZ-TH/07-21\\
0807.2148\\
July 2008
\end{flushright}

\begin{center}
{\Large\bf Heavy baryon properties with\\[12pt]
NLO accuracy in perturbative QCD} \\[1truecm]
{\large S.~Groote,$^{1,2}$ J.G.~K\"orner$^1$ and
  A.A.~Pivovarov$^{1,3}$}\\[.7cm]
$^1$ Institut f\"ur Physik der Johannes-Gutenberg-Universit\"at,\\
  Staudinger Weg 7, 55099 Mainz, Germany\\[.5truecm]
$^2$ Loodus- ja Tehnoloogiateaduskond, Tartu \"Ulikool,
  T\"ahe 4, 51010 Tartu, Estonia\\[.3truecm] 
$^3$ Institute for Nuclear Research of the\\
Russian Academy of Sciences, Moscow 117312, Russia
\vspace{1truecm}
\end{center}

\begin{abstract}
We present an analysis of the static properties of heavy baryons at
next-to-leading order in the perurbative expansion of QCD. We obtain
analytical next-to-leading order three-loop results for the two-point
correlators of baryonic currents with one finite mass quark field for a
variety of quantum numbers of the baryonic currents.  We consider both the
massless limit and the HQET limit of the correlator as special cases of the
general finite mass formula and find agreement with previous results. We
present closed form expressions for the moments of the spectral density. We
determine the residues of physical baryon states using sum rule techniques.
\end{abstract}

\newpage

\tableofcontents

\newpage

\section{Introduction}
Baryons form a rich family of particles which have been experimentally studied
with high accuracy~\cite{PDG}. With the advent of new accelerators and
detectors many properties of baryons containing a heavy quark have been
experimentally measured in recent years~\cite{PDG}. A theoretical analysis of
these experimental data gives a great deal of information about the structure
of QCD and the numerical values of its parameters. The hypothetical limit
$N_c\to\infty$ for the number $N_c$ of quark and gluon colours in the symmetry
group $SU(N_c)$ was especially successful for baryons~\cite{Witten:1979kh}. 
The analysis of this limit is a very powerful tool for the investigation of
general
properties of gauge interactions. The information about the spectrum of 
baryons is contained in the correlator of two baryonic currents and the
spectral density associated with it. Within the operator product expansion and
to leading order in perturbative QCD the correlator is given by a product of
$N_c$ fermionic propagators. The diagrams of this topology have been studied
in some detail~\cite{Berends:1997vk,sunrise,thresh}. They are rather
frequently used in phenomenological applications~\cite{sunapp,pivbar,pivan}.
In general, these diagrams represent the leading order of the perturbative
expansion for the relevant correlation functions. In some cases, though
(especially for the gluon current correlators ~\cite{Kataev:1982gr}), they
first appear at next-to-leading order. Complete calculations beyond the
leading order have not been done for many interesting cases. In this paper we
fill this gap.

We report on the results of calculating the $\alpha_s$ corrections to the
correlator of two baryonic currents with one finite mass quark and two
massless quarks. We present analytical results and discuss the magnitude of
the $\alpha_s$ corrections for the physically interesting cases. The high 
energy (massless quarks) and near-threshold (Heavy Quark Effective Theory,
HQET)~\cite{Neubert:1993mb,Mannel:1990vg,Mannel:1991mc,Korner:1991kf} limits
are obtained from our general results as special cases. For these we find 
agreement with previous results in the literature. We present analytical
results for the moments of the spectral density associated with correlators of
baryonic currents.

We briefly discuss the impact of our new results for the baryonic correlators
on the phenomenology of baryons. However, the main aim of this paper is to
present the results of the perturbative calculations in some detail and to 
show how they
have been arrived at. The new technically demanding feature of our calculation
is the presence of a finite-mass quark in the correlator which is needed for
baryons containing a heavy charm or bottom quark. For the $b$-quark the
accuracy of the HQET approximation is rather good. The exact result will help
in controlling the precision of the approximation. For the $c$-quark, however,
the accuracy of the near threshold approximation is insufficient for physical
applications and the use of the exact formulas is unavoidable.

The massless case has been known since long ago~\cite{pivbar,pivan} and serves
as a test of the massless limit of our results. We mention that the mesonic
analogue of our baryonic calculation with one finite mass quark and one
massless antiquark was completed some time ago~\cite{Generalis} and has
subsequently provided a rich source of inspiration for many applications in
meson physics. 

Some of the techniques used in this paper have already been usefully employed 
in the analysis of perturbative corrections to sum rules involving pentaquark
states~\cite{Groote:2006sy}.

\newpage

\section{Generalities}
In this section we present our choice of interpolating currents for baryons
containing one heavy quark. We also introduce two-point correlation functions
as the principal tool in our analysis of the static properties of heavy
baryons. Finally, we give an outline of the techniques that were used in our
calculations.
 
\subsection{Choice of currents and correlators}
A generic lowest dimensional baryonic current has the form
\begin{equation}\label{cur0}
j=\epsilon^{abc}(u_b^TC\Gamma d_c)\Gamma'\Psi_a.
\end{equation}
The current (\ref{cur0}) refers to a baryon with three valence quarks and
no gluonic fields and no derivative couplings. $\Psi$ is a finite mass quark
field with the mass parameter $m$ and $u$ and $d$ are massless quark fields.
$C$ is the charge conjugation matrix, $\epsilon^{abc}$ is the totally
antisymmetric tensor and $a,b,c$ are colour indices of the $SU(3)$ colour
group. $\Gamma$ and $\Gamma'$ stand for Dirac matrices or strings of Dirac
matrices where possible Lorentz indices on $\Gamma$ and $\Gamma'$ such as
in $\gamma^{\mu}$ or $\sigma^{\mu \nu}$ are suppressed. In much the same way we
have suppressed a possible Lorentz index on $\Psi_a$ which is needed later on
in the discussion of the spin $3/2$ field. For $\Gamma=1$, 
$\Gamma'=\gamma_5$ the interpolating current has the
quantum numbers of a $J^P=1/2^+$ baryon. Other baryonic currents with any
given specified quantum numbers are obtained from the current in
Eq.~(\ref{cur0}) by using the appropriate Dirac matrices or strings of Dirac
matrices. We first consider the simplest
case and take $\Gamma=1$, $\Gamma'=1$ corresponding to an
interpolating current with quantum numbers $1/2^-$. This allows us to
explain our techniques and to demonstrate the idiosyncratic features of the 
calculation. Later in the text we will introduce more general interpolating 
currents and discuss calculational differences in comparison to those in the 
simplest case.

The correlator of two baryonic currents can be
expanded into a basis of invariant functions. The form of this expansion
depends on the Dirac matrices employed. In the simplest case 
$\Gamma=1$, $\Gamma'=1$ there are only
two invariant functions $\Pi^q(q^2)$ and $\Pi^m(q^2)$ in the expansion which 
are defined through
\begin{equation}\label{def00gen}
i\int\langle T\{j(x)\bar j(0)\}\rangle e^{iqx}dx=m\Pi^m(q^2)+\slq\Pi^q(q^2).
\end{equation}
In the following we shall refer to the two contributions on the r.h.s. of
(\ref{def00gen}) as the mass and the momentum term, respectively. Note that
each of the replacements 
$\Gamma\to\Gamma\gamma_5$ and $\Gamma'\to\Gamma'\gamma_5$ leads to the change 
$\Pi^q(q^2)\to-\Pi^q(q^2)$.

\subsection{Basic techniques}
The generic correlation function $\Pi(q^2)$ has a dispersion representation
\begin{equation}\label{eq:dispbasic}
\Pi(q^2)=\int_{m^2}^\infty\frac{\rho(s)ds}{s-q^2}+\mbox{subtractions}
\end{equation}
through its discontinuity $\rho(s)$ on the physical cut $s>m^2$,
\begin{equation}
\rho(s)=\frac1{2\pi i}\Big(\Pi(s+i0)-\Pi(s-i0)\Big)\,.
\end{equation}
The discontinuity can also be written as the imaginary part of the correlation
function if the phases are properly chosen,
\begin{equation}
\rho(s)=\frac1\pi\mbox{Im}\Pi(s+i0).
\end{equation}
The expression for the discontinuity or spectral density $\rho(s)$ is simpler 
than the expression for the correlation function $\Pi(q^2)$ itself. The 
knowledge of $\rho(s)$ suffices for physical purposes and allows one to 
recover the whole function $\Pi(q^2)$ through a one-dimensional integral with 
a simple weight function as given in Eq.~(\ref{eq:dispbasic}). For this reason
we concentrate on calculating the spectral density $\rho(s)$.

The general strategy is rather straightforward. The main part of the
calculation is done by using a symbolic manipulation program. One reduces all
integrals to some basic master integrals and then one puts them together again
to get the result for a particular correlator with any given quantum numbers.
This program has been explicitly realized in our evaluation.

The topology of the NLO diagrams is such that at least one line connecting the
initial and final points of the diagram is free. If this line is the massive
one, the remaining part of the diagram consists of massless lines and can be
integrated analytically. Adding the massive line leads to a one-dimensional
integration which can be done analytically.

If the massive line is part of the radiative corrections, the basic quantity 
is a NLO two-point correlator of the meson type with one heavy and one light
quark. The spectral density for this NLO correlator is known to be computable.
The fact that we only have one finite mass and a special topology of diagrams
in the baryon sector therefore makes the analytical computation feasible. Note
that the computation with two different finite masses can still be done
but requires a numerical calculation while some limiting cases such as the
small mass ratio limit can still be done analytically.

In order to explain our main tools, we consider the correlation function of
the baryonic current, which, up to NLO, can be written as   
\begin{equation}
\Pi(q^2)=\int {dk} \Pi_2(q-k)\Pi_1(k)
\end{equation}
where $\Pi_{1,2}(k)$ are one- and two-line correlators. If the one-line
correlator is massive, the massless two-line correlator can be explicitly
integrated and we are left with an integral in $D$-dimensional space-time
given by
\begin{equation}
V(\alpha,\beta)=\int\frac{d^Dk}{(m^2-k^2)^\alpha(-(q-k)^2)^\beta}.
\end{equation}
This integral can be expressed through hypergeometric functions and is
therefore completely known. If the one-line correlator is massless, the
spectral density reads
\begin{equation}
\rho(s)=\int_{m^2}^{s}ds'\rho_2(s')(1-s'/s).
\end{equation}
It is not difficult to obtain $\rho(s)$ since $\rho_2(s)$ is known from 
mesonic type calculations.

All in all the calculation includes no unknown elements in the sense that all 
necessary blocks (prototypes or masters) are known to be calculable
analytically. The main problem is the reduction of the initial diagrams to
prototypes and the assembly of the final results from these building blocks.
This has been done using the computer.

In the next section we present the calculation and results for the lowest spin
baryons which means that the Dirac matrices in the interpolating currents are
just unity or $\gamma_5$.

\section{Lowest spin baryons}
In this section we present the results for the simplest choice of the Dirac
structure of interpolating baryonic currents. We take $\Gamma=1$ and
$\Gamma'=\gamma_5$ which corresponds to an interpolating current with quantum
numbers $J^P=1/2^+$. We can in fact omit $\gamma_5$ in the process of the 
calculation because the effect of $\gamma_5$ can later on be easily 
accounted for by a simple multiplication in Dirac space. In this case the 
basic baryonic current has the form
\begin{equation}\label{cur}
j=\epsilon^{abc}(u_b^TCd_c)\Psi_a.
\end{equation}
For this scalar case with $\Gamma=\Gamma'=1$, the results for the invariant 
functions $\Pi^m(q^2)$ and $\Pi^q(q^2)$ in Eq.~(\ref{def00gen}) have already 
been presented in Refs.~\cite{masspart,mompart}. The invariant function 
$\Pi^\alpha(q^2)$ with $\alpha\in\{q,m\}$ can be represented compactly via 
the dispersion relation 
\begin{equation}
\Pi^\alpha(q^2)=\int_{m^2}^\infty\frac{\rho^\alpha(s)ds}{s-q^2}
\end{equation}
where $\rho^\alpha(s)$ is the spectral density. All quantities are understood
to be appropriately regularized. Since the spectral density is the main object
of interest for phenomenological applications, we limit our subsequent
discussion to the spectral density
\begin{equation}
\rho^\alpha(s)=\frac{s^2}{128\pi^4}\left\{\rho_0^\alpha(s)
  \left(1+\frac{\alpha_s}\pi\ln\pfrac{\mu^2}{m^2}\right)
  +\frac{\alpha_s}\pi\rho_1^\alpha(s)\right\},
\end{equation}
where $\mu$ is the renormalization scale parameter, $m$ is the pole mass of
the heavy quark (see e.g.\ Ref.~\cite{Tarrach:1980up}), and
$\alpha_s=\alpha_s(\mu)$.

\subsection{LO analytical results}
\begin{figure}[t]\begin{center}
\vbox{
\epsfig{figure=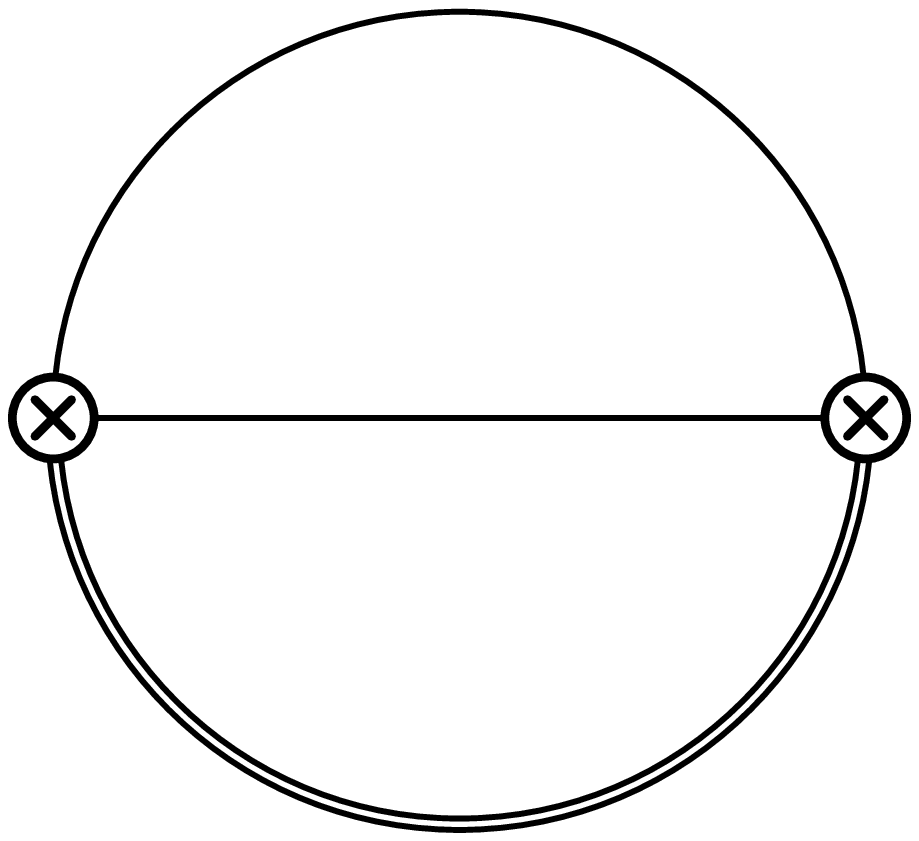, scale=0.3}
\epsfig{figure=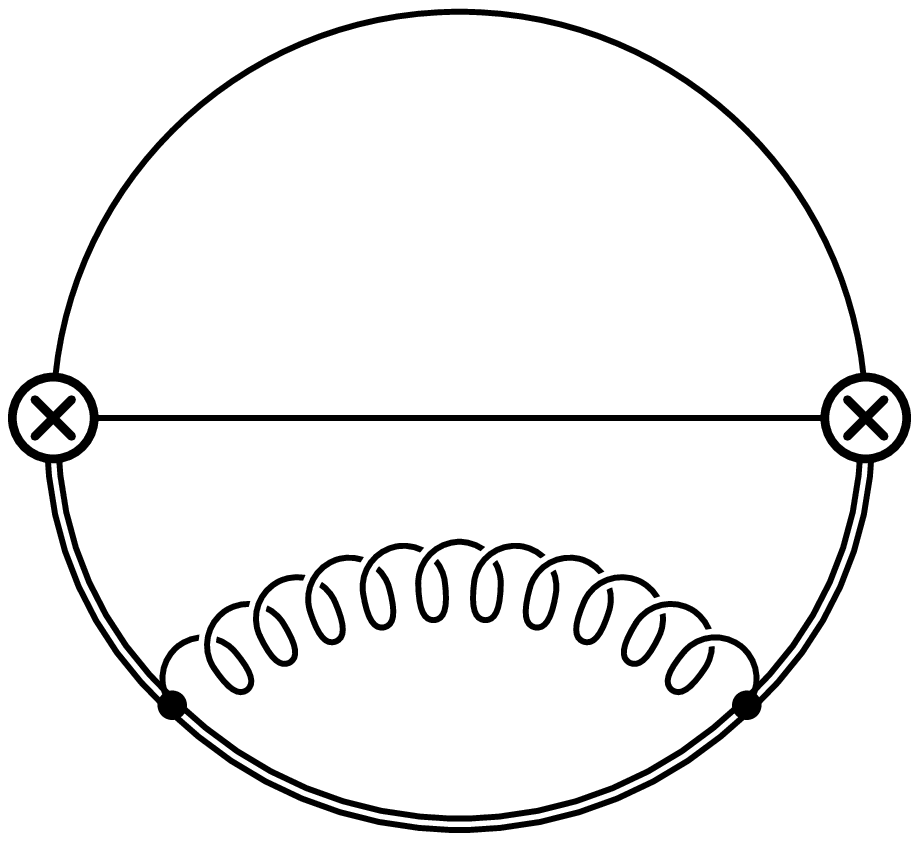, scale=0.3}
\epsfig{figure=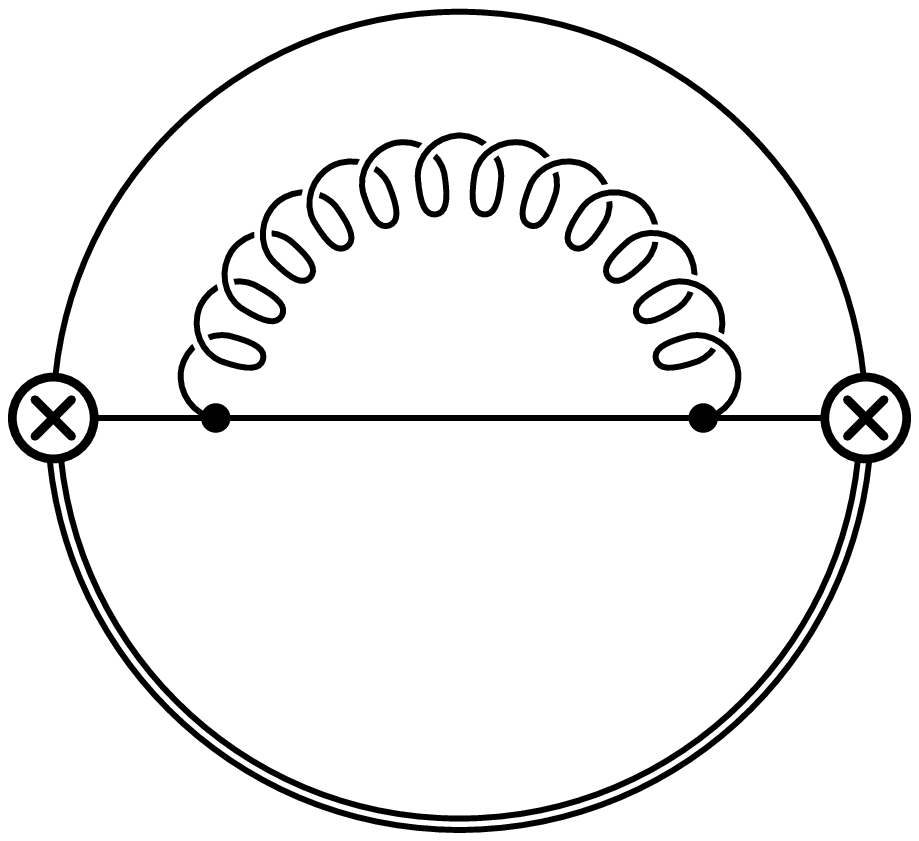, scale=0.3}
\epsfig{figure=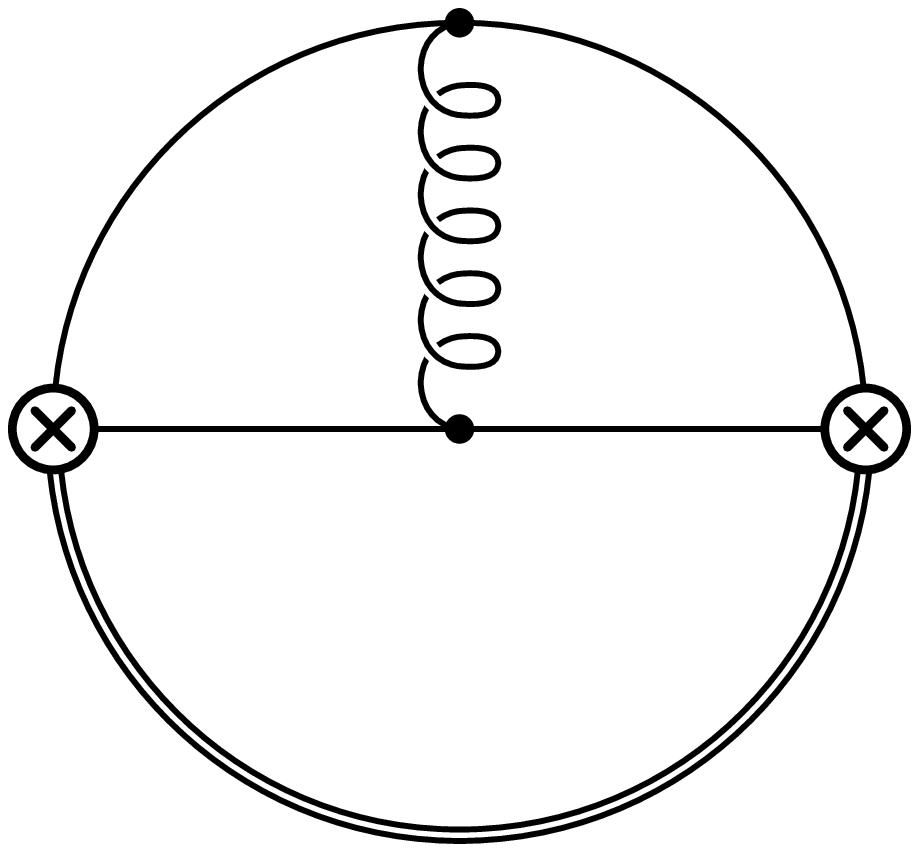, scale=0.3}
\epsfig{figure=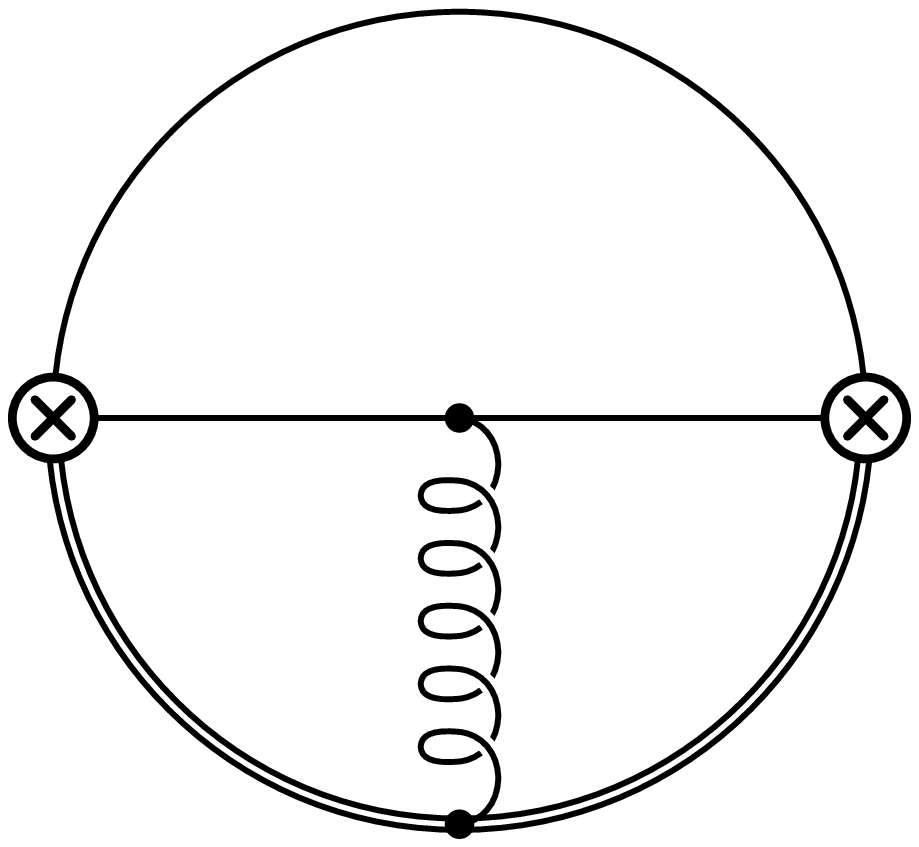, scale=0.3}\\
  (a1)\kern56pt(b11)\kern56pt(b21)\kern56pt(c11)\kern56pt(c21)}
\caption{\label{fig1} Two-loop (a1) and three-loop (b11--c21) topologies with
  one external momentum. Heavy lines represent the heavy quark and light lines
  massless quarks.}
\end{center}\end{figure}

The leading order two-loop diagram is shown in Fig.~\ref{fig1}(a). Note that
this topology coincides with what is referred to as sunrise-type diagrams for
which a general evaluation method (with arbitrary masses) has been developed 
in Refs.~\cite{sunrise,Groote:2004qq}. Sunrise-type diagrams can be calculated
by a variety of methods. In this paper we apply the configuration space
technique in a straightforward manner. The result reads
\begin{equation}\label{lead0m}
\rho_0^m(s)=1+9z-9z^2-z^3+6z(1+z)\ln z 
\end{equation}
\begin{equation}\label{lead0q}
\rho_0^q(s)=\frac14-2z+2z^3-\frac14z^4-3z^2\ln z
\end{equation}
with $z=m^2/s$. 

\subsection{NLO analytical results}
The contributing three-loop diagrams are shown in Figs.~\ref{fig1}(b11) to
(c21). They have been evaluated using the advanced algebraic methods for
multi-loop calculations along the lines decribed in Refs.~\cite{sunrise}.
The result can be obtained analytically. In the $\msbar$-subtraction scheme
one has \cite{masspart,mompart}
\begin{eqnarray}\label{corr1m}
\lefteqn{\rho_1^m(s)\ =\ 9+\frac{665}9z-\frac{665}9z^2-9z^3
  -\left(\frac{58}9+42z-42z^2-\frac{58}9z^3\right)\ln(1-z)}\nonumber\\&&
  +\left(2+\frac{154}3z-\frac{22}3z^2-\frac{58}9z^3\right)\ln z
  +\frac83\left(1+9z-9z^2-z^3\right)\left(\Li_2(z)
  +\frac12\ln(1-z)\ln z\right)\nonumber\\&&\kern-18pt
  +z\left(24+36z+\frac43z^2\right)\left(\Li_2(z)-\zeta(2)+\frac12\ln^2z\right)
  +24z(1+z)\left(\Li_3(z)-\zeta(3)-\frac13\Li_2(z)\ln z\right),\nonumber\\\\
\label{corr1q}
\lefteqn{\rho_1^q(s)\ =\ 
  \frac{71}{48}-\frac{565}{36}z-\frac78z^2+\frac{625}{36}z^3-\frac{109}{48}z^4
  -\frac1{36}\left(\frac{49}{36}-\frac{116}9z+\frac{116}9z^3-\frac{49}{36}z^4
  \right)\ln(1-z)}\nonumber\\&&\kern-18pt
  +\left(\frac14-\frac{17}3z-11z^2+\frac{113}9z^3-\frac{49}{36}z^4\right)\ln z
  +\frac23\left(1-8z+8z^3-z^4\right)\left(\Li_2(z)+\frac12\ln(1-z)\ln z\right)
  \nonumber\\&&\kern-12pt
  -\frac13z^2\left(54+8z-z^2\right)\left(\Li_2(z)-\zeta(2)+\frac12\ln^2z\right)
  -12z^2\left(\Li_3(z)-\zeta(3)-\frac13\Li_2(z)\ln(z)\right)
\end{eqnarray}
where $z=m^2/s$ and $\Li_n(z)$ are polylogarithms 
\begin{equation}
\Li_n(z)=\sum_{k=1}^\infty\frac{z^k}{k^n},\qquad
\Li_n(1)=\sum_{k=1}^\infty\frac1{k^n}=\zeta(n).
\end{equation}
$\zeta(n)$ is Riemann's zeta function. Note that in the physical region we
have $z<1$. Therefore, no analytic continuation is required. Note the
repeated appearance of particular combinations of polylogarithms and
logarithms in the form
\begin{eqnarray}
\Li_3(z)-\zeta(3)-\frac13\Li_2(z)\ln z
  &=&\frac13\int_1^z\frac{dz'}{z'}\left(2\Li_2(z')+\ln(1-z')\ln z'\right),
  \nonumber\\
\Li_2(z)-\zeta(2)+\frac12\ln^2z
  &=&\int_1^z\frac{dz'}{z'}\left(\ln z'-\ln(1-z')\right)
  \quad\mbox{and}\nonumber\\
\Li_2(z)+\frac12\ln(1-z)\ln z
  &=&-\frac12\int_0^zdz'\left(\frac{\ln z'}{1-z'}+\frac{\ln(1-z')}{z'}\right)
\end{eqnarray}
appear as a consequence of the integration of the two-line spectral functions.

\section{General baryon case}
The baryonic current defined in Eq.~(\ref{cur0}) has the most general
structure concerning possible choices for the Dirac matrices $\Gamma$ and
$\Gamma'$. However, not all of this structure has to be kept in order to
calculate the different invariant functions of the correlator. In
four-dimensional space-time where the initial (bare) current is defined, Fierz
rearrangement can always be used to take the heavy spinor out of the trace in
the correlator. When taken out of the trace together with the heavy spinor,
the matrix $\Gamma'$ can be considered to be an overall factor that has no
effect on the calculation. Indeed,
\begin{equation}\label{curGenSim}
j=\epsilon^{abc}(u_b^TC\Gamma d_c)\Gamma'\Psi_a=\Gamma'j_\Gamma
\end{equation}
where $j_\Gamma$ is the current for $\Gamma'=1$,
\begin{equation}\label{curGenSim1}
j_\Gamma=\epsilon^{abc}(u_b^TC\Gamma d_c)\Psi_a.
\end{equation}
The general case is recovered as
\begin{equation}\label{def0011}
\Pi(q^2)=i\int\langle T\{j(x)\bar j(0)\}\rangle e^{iqx}dx
  =\Gamma'\Pi_\Gamma(q^2)\bar\Gamma'
\end{equation}
with $\bar\Gamma=\gamma^0\Gamma^\dagger\gamma^0$ and
\begin{equation}\label{def001122}
\Pi_\Gamma(q^2)=i\int\langle Tj_\Gamma(x)\bar j_\Gamma(0)\rangle e^{iqx}dx.
\end{equation}
Because the general result can easily be recovered, we can limit ourselves to
the case $\Gamma'=1$. The calculation of the correlator has to be done with
its full dependence on the matrix $\Gamma$. We obtain the general expression
for the correlator in the form 
\begin{equation}
\Pi_\Gamma(q^2)=\sum_{i=1}^6A_i(q^2)\tr_i(\Gamma,q^2)
\end{equation}
where
\begin{eqnarray}\label{traces}
\tr_1(\Gamma,q^2)&=&\Tr(\Gamma\slq\bar\Gamma\slq)m/q^2\quad\ \
\tr_2(\Gamma,q^2)\ =\ \Tr(\Gamma\gamma_\alpha\bar\Gamma\gamma^\alpha)m
  \nonumber\\[3pt]
\tr_3(\Gamma,q^2)&=&\Tr(\Gamma\slq\bar\Gamma\slq)\slq/q^2\qquad
\tr_4(\Gamma,q^2)\ =\ \Tr(\Gamma\slq\bar\Gamma\gamma_\alpha)\gamma^\alpha
  \nonumber\\[3pt]
\tr_5(\Gamma,q^2)&=&\Tr(\Gamma\gamma_\alpha\bar\Gamma\slq)\gamma^\alpha\qquad\
\tr_6(\Gamma,q^2)\ =\ \Tr(\Gamma\gamma_\alpha\bar\Gamma\gamma^\alpha)\slq.
\end{eqnarray}
The trace in (\ref{traces}) is to be taken only with respect to the 
$\gamma$-string in the round brackets. We have again omitted possible
Lorentz indices on $\Gamma$ such as appear in the next example. 

As an example, let us exhibit the structure of the expressions
for the interesting and important case of the ``vector'' current
\begin{equation}\label{veccur}
j^\mu=\epsilon^{abc}(u_b^TC\gamma^\mu d_c)\Psi_a.
\end{equation}
In the above notation this means that $\Gamma'=1$.
The expansion of the correlator reads
\begin{equation}\label{vectorcorrel}
\Pi^{\mu\nu}(q^2)=i\int\langle Tj^\mu(x)\bar j^\nu(0)\rangle e^{iqx}dx.
\end{equation}
The correlator can be expanded along a set of ten covariants. The expansion 
reads
\begin{eqnarray}\label{vectordecomp}
\Pi^{\mu\nu}(q^2)&=&m\left(A_1^mq^\mu q^\nu+A_2^mq^2g^{\mu\nu}
  +A_3^m\slq q^\mu\gamma^\nu+A_4^m\slq\gamma^\mu q^\nu
  +A_5^mq^2\gamma^\mu\gamma^\nu\right)/q^2\nonumber\\&&
  +\slq\left(A_1^qq^\mu q^\nu+A_2^qq^2g^{\mu\nu}
  +A_3^q\slq q^\mu\gamma^\nu+A_4^q\slq\gamma^\mu q^\nu
  +A_5^qq^2\gamma^\mu\gamma^\nu\right)/q^2.
\end{eqnarray}
All invariant amplitudes $A_i^\alpha$ have been calculated. The above 
expansion of the vector correlator is only one of several possible expansions.
By making use of the symmetry properties of the result we shall later on use
a different expansion in terms of nine covariants whose mixing property is
simpler.

\subsection{LO analytical results}
Differing from our earlier calculation \cite{masspart,mompart}, where the mass
and momentum parts were calculated separately, we have now learned to
calculate the mass and momentum parts in one go. The reason is that no
explicit traces have to be taken when extracting the mass or momentum part.
Instead, the traces are kept to the very end. It is then not difficult to
interprete one part of the expression as the mass part and the other part as
the momentum part depending on the occurence or absence of an explicit factor
of $m$. The result for the leading order diagram (a1) in $D=4-2\eps$ space-time
dimensions reads
\begin{eqnarray}\label{lores}
\rho_{a1}(s)&=&\frac{(D-2)G(1,1)N_c!}{16(4\pi)^D(D-1)^2}s^{D-2}
  \hat\rho_{a1}(m^2/s)\qquad\mbox{where}\nonumber\\[3pt]
\hat\rho_{a1}(z)&=&\sum_{i=1}^6\hat\rho_{a1}^i(z)\tr_i(\Gamma,s).
\end{eqnarray}
Here $G(1,1)=G/\eps$ and $G=\Gamma(1+\eps)\Gamma(1-\eps)^2/\Gamma(2-2\eps)$.
Note that we use hatted spectral functions whenever we present them as a
function of $z=m^2/s$. The corrections are
\begin{eqnarray}\label{leadbase}
\hat\rho_{a1}^1(z)&=&D\Bigg\{\hat\rho_V(1,\eps-2;z)+(1-z)^2\hat\rho_V(1,\eps;z)
  -2\left(\frac{D-2}D-z\right)\hat\rho_V(1,\eps-1;z)\Bigg\},\nonumber\\
\hat\rho_{a1}^2(z)&=&-\Bigg\{\hat\rho_V(1,\eps-2;z)+(1-z)^2\hat\rho_V(1,\eps;z)
  +2\left(\frac{3D-4}{D-2}+z\right)\hat\rho_V(1,\eps-1;z)\Bigg\},\nonumber\\
\hat\rho_{a1}^3(z)&=&\frac{D+2}2\Bigg\{\hat\rho_V(1,\eps-3;z)
  +\left(\frac{D-2}{D+2}+z\right)(1-z)^2\hat\rho_V(1,\eps;z)
  \,+\nonumber\\&&
  +\left(\frac{6-D}{D+2}+3z\right)\hat\rho_V(1,\eps-2;z)
  +\left(\frac{2-D}{D+2}-\frac{2Dz}{D+2}+3z^2\right)
  \hat\rho_V(1,\eps-1;z)\Bigg\},\nonumber\\
\hat\rho_{a1}^4(z)&=&-\frac12\Bigg\{\hat\rho_V(1,\eps-3;z)
  -(1-z)^3\hat\rho_V(1,\eps;z)\,+\nonumber\\&&\qquad
  +(1+3z)\hat\rho_V(1,\eps-2;z)
  -(1-z)(1+3z)\hat\rho_V(1,\eps-1;z)\Bigg\},\nonumber\\
\hat\rho_{a1}^5(z)&=&-\frac12\Bigg\{\hat\rho_V(1,\eps-3;z)
  -(1-z)^3\hat\rho_V(1,\eps;z)\,+\nonumber\\&&\qquad
  +(1+3z)\hat\rho_V(1,\eps-2;z)
  -(1-z)(1+3z)\hat\rho_V(1,\eps-1;z)\Bigg\},\nonumber\\
\hat\rho_{a1}^6(z)&=&-\frac12\Bigg\{\hat\rho_V(1,\eps-3;z)
  +(1-z)^2(1+z)\hat\rho_V(1,\eps;z)\,+\nonumber\\&&\qquad
  +\left(\frac{7D-10}{D-2}+3z\right)\hat\rho_V(1,\eps-2;z)
  \,+\nonumber\\&&\qquad\qquad
  +\left(\frac{7D-10}{D-2}+2\frac{3D-4}{D-2}z+3z^2\right)
  \hat\rho_V(1,\eps-1;z)\Bigg\}.
\end{eqnarray}
For later reference we also need starred elements $\hat\rho_{a1}^{1*}(z)$,
$\hat\rho_{a1}^{2*}(z)$, $\hat\rho_{a1}^{3*}(z)$, $\hat\rho_{a1}^{4*}(z)$,
$\hat\rho_{a1}^{5*}(z)$, and $\hat\rho_{a1}^{6*}(z)$. These can be obtained
from the corresponding unstarred elements by replacing $\eps$ in the argument
of the spectral functions $\hat\rho_V$ by $2\eps$. The basic spectral
functions $\tilde\rho_V$ are given~by
\begin{equation}\label{rhovdef}
\hat\rho_V(n_1,n_2;z)=\frac1{\Gamma(n_1)\Gamma(n_2)\Gamma(1+D/2-n_1-n_2)}
\int_z^1(1-x)^{D/2-n_2-1}x^{n_2-1}(x-z)^{D/2-n_1-n_2}dx
\end{equation}
where $\Gamma(z)$ is Euler's gamma function. Note that the singularity given
by the factor $G(1,1)$ in Eq.~(\ref{lores}) cancels against the singularity
of the second gamma functions in the denominator if the appropriate arguments
$n_2=\eps-n$ (or $n_2=2\eps-n$ in case of the starred elements) with $n\ge 0$
are used. In the limit $D=4$ we find
\begin{eqnarray}\label{LOfullres}
\rho_{a1}(s)&=&\frac{s^2}{512\pi^4}\sum_{i=1}^6\hat\rho_{a1}^{i0}(m^2/s)
  \tr_i(\Gamma,s)
\end{eqnarray}
where
\begin{eqnarray}\label{LOfullres:terms}
\hat\rho_{a1}^{10}(z)&=&\frac12+\frac53z-3z^2+z^3-\frac16z^4+2z\ln z\nonumber\\
\hat\rho_{a1}^{20}(z)&=&\frac18+\frac{11}6z-\frac32z^2-\frac12z^3
  +\frac1{24}z^4+\left(z+\frac32z^2\right)\ln z\nonumber\\
\hat\rho_{a1}^{30}(z)&=&\frac1{10}-\frac12z+z^2-z^3+\frac12z^4-\frac1{10}z^5
  \nonumber\\
\hat\rho_{a1}^{40}(z)&=&\frac1{40}-\frac14z-\frac16z^2+\frac12z^3-\frac18z^4
  +\frac1{60}z^5-\frac12z^2\ln z\nonumber\\
\hat\rho_{a1}^{50}(z)&=&\frac1{40}-\frac14z-\frac16z^2+\frac12z^3-\frac18z^4
  +\frac1{60}z^5-\frac12z^2\ln z\nonumber\\
\hat\rho_{a1}^{60}(z)&=&\frac1{40}-\frac14z-\frac16z^2+\frac12z^3-\frac18z^4
  +\frac1{60}z^5-\frac12z^2\ln z
\end{eqnarray}
Note that $\hat\rho_{a1}^{40}(z)=\hat\rho_{a1}^{50}(z)=\hat\rho_{a1}^{60}(z)$.
Eqs.~(\ref{LOfullres}) and~(\ref{LOfullres:terms}) give the full answer for
the leading order contribution to the baryonic correlators for all possible
configurations of Dirac gamma matrices. In this sense this completes the 
leading order calculation of correlators for baryons with any quantum numbers
as long as there are no derivative couplings in the interpolating currents.  

\subsection{NLO contributions}
The NLO contributions result from four different diagrams. In the calculation
of these diagrams we have used different techniques depending on their
topologies and on the location of the massive line (cf.\ Fig.~\ref{fig1}).
\begin{itemize}
\item The {\em NLO light contributions\/} result from two diagrams, the self
energy correction of one of the massless lines (b21) and the diagram with
gluon exchange between two massless lines (c11) which we term ``light
fish''. The technique of calculating these diagrams
consists in first analytically calculating the massless part and then adding
the massive fermion line. Advantage is taken of the fact that massless
two-loop diagrams are explicitly calculable. In this case one cannot avail
of a convolution of spectral functions.
\item The {\em massive line self energy contribution\/} (b11) was calculated
again by using the explicit evaluation of the massless part and the dispersion
relation for the mass operator of the massive quark at leading order. Here
we define the heavy quark mass through its pole mass which is
most convenient for the calculation using cuts.
\item The most demanding {\em semi-massive fish diagram\/} (c21) was
calculated by making full use of the decomposition into prototypes and the
convolution of spectral functions. For this diagram the use of symbolic
manipulation programs is indispensable as the number of terms in
intermediate expressions involving different structures are quite large.
\end{itemize}
These three main parts constitute the whole calculation.

\subsection{Light contributions (b21 and c11)}
The NLO light contributions result from two diagrams, the self energy
correction of one of the massless lines (b21) and the diagram with gluon
exchange between the two massless lines (c11), called ``light fish''. More
details on the calculation of these two contributions are found in Appendix~A.
It turns out that the dominant singular part is proportional to the leading
order contribution. The result for the spectral density reads
\begin{equation}\label{lightcon}
\rho_{\rm light}(s)=\frac{\alpha_sN_c!C_F}{4\pi}\left\{
  \left(\frac{B_0}{\eps^2}+\frac{B_1}\eps+B_2\right)\rho^*_{a1}(s)
  +\left(\frac{B'_1}\eps+B'_2\right)\rho_{a1}^{\prime*}(s)\right\}
\end{equation}
where
\begin{eqnarray}
\rho^*_{a1}(s)&=&\frac{(D-2)G^2s^{3D/2-4}}{16(4\pi)^D(D-1)^2}
  \sum_{i=1}^6\hat\rho_{a1}^{*i}(m^2/s)\tr_i(\Gamma,s),\nonumber\\[3pt]
\rho_{a1}^{\prime*}(s)&=&\frac{G^2s^{3D/2-4}}{2(4\pi)^DD}
  \Big(\hat\rho_{a1}^{m*}(m^2/s)\tr_2(\Gamma,s)
  +\hat\rho_{a1}^{q*}(m^2/s)\tr_6(\Gamma,s)\Big)
\end{eqnarray}
and where $\hat\rho_{a1}^{*i}(z)$ are the starred elements introduced after
Eq.~(\ref{leadbase}). The spectral functions
\begin{eqnarray}
\hat\rho_{a1}^{m*}(z)&=&-\hat\rho_V(1,2\eps-1;z),\nonumber\\
\hat\rho_{a1}^{q*}(z)&=&-\frac12\left\{(1+z)\hat\rho_V(1,2\eps-1;z)
  +\hat\rho_V(1,2\eps-2;z)\right\}
\end{eqnarray}
are known from the scalar calculation, and
\begin{eqnarray}\label{bcoeffs}
\frac{B_0}{\eps^2}+\frac{B_1}\eps+B_2
  &=&\left(\frac1{\eps^2}-\frac1{6\eps}+\frac{17}{12}\right)
  -\frac{C_B}{C_F}\left[\frac{c_\Gamma^2}4\left(\frac1{\eps^2}+\frac1{3\eps}
  -\frac16\right)+3\left(\frac1\eps+\frac{19}6-4\zeta(3)\right)\right],
  \nonumber\\
\frac{B'_1}\eps+B'_2&=&\frac23\left(\frac1\eps+2\right)
  -\frac{C_B}{C_F}\left[\frac{c_\Gamma^2}6\left(\frac1\eps+\frac52\right)+1
  \right]
\end{eqnarray}
where $C_F=(N_c^2-1)/2N_c=4/3$, $C_B=(N_c+1)/2N_c=2/3$ for $N_c=3$ colours.
The occurence of $C_F$ in the denominator of the second parts in
Eq.~(\ref{bcoeffs}) results from the fact that $C_F$ is factored out in
Eq.~(\ref{lightcon}). The parameter $c_\Gamma$ is defined by
\begin{equation}
\label{cgamma}
\gamma_\alpha\bar\Gamma\gamma^\alpha=c_\Gamma\bar\Gamma,\qquad
c_\Gamma=s_\Gamma(D-2r_\Gamma)
\end{equation}
and can be expressed by the signature $s_\Gamma=\pm1$ of the matrix according
to whether one has an even or odd number of $\gamma$--matrices (including 
$\gamma_5$) and the
number $r_\Gamma$ of Dirac matrices other than $\gamma_5$. The fact that the
most singular term is proportional to the leading order contribution allows
one to extract a common renormalization factor. The result reads
\begin{eqnarray}
\rho_l(s)&=&\frac{N_c!s^2}{18(4\pi)^4}\Bigg\{
  \left(1+\frac{\alpha_sC_F}{4\pi\eps}C^{r(0)}\right)\hat g_1(m^2/s)
  \nonumber\\&&\qquad
  +\frac{\alpha_sC_F}{4\pi}\left(C^{r(1)}\hat g_1^{(0)}(m^2/s)
  +C^{r(0)}\hat g_{21}^{(1)}(m^2/s)
  +C^{r\prime(0)}\hat g_2^{\prime(0)}(m^2/s)\right)
  \Bigg\}.
\end{eqnarray}
where
\begin{equation}
C^{r(0)}=2B_0G,\qquad C^{r(1)}=2\left(B_0\ln\pfrac{\mu^2}s+B_0+B_1\right)G,
\qquad C^{r\prime(0)}=2B'_1G.
\end{equation}
The definition of the functions $\hat g_i^\alpha(z)$ can be found in
Appendix~A.

\subsection{The massive line self energy contribution (b11)}
We divide the leading order contribution in Eq.~(\ref{lores}) into a mass
part
\begin{eqnarray}
\rho_{a1}^m(s)&=&\frac{(D-2)G(1,1)N_c!}{16(4\pi)^D(D-1)^2}s^{D-2}
  \hat\rho_{a1}^m(m^2/s)\,,\qquad\mbox{where}\nonumber\\[3pt]
\hat\rho_{a1}^m(m^2/s)&=&\sum_{i=1}^2\hat\rho_{a1}^i(m^2/s)\tr_i(\Gamma,s)\,,
\end{eqnarray}
and a momentum part
\begin{eqnarray}
\rho_{a1}^q(s)&=&\frac{(D-2)G(1,1)N_c!}{16(4\pi)^D(D-1)^2}s^{D-2}
  \hat\rho_{a1}^q(m^2/s)\,,\qquad\mbox{where}\nonumber\\[3pt]
\hat\rho_{a1}^q(m^2/s)&=&\sum_{i=3}^6\hat\rho_{a1}^i(m^2/s)\tr_i(\Gamma,s)\,.
\end{eqnarray}
The result for the self energy correction of the massive line is obtained by
a convolution of these leading order contributions. One obtains
$\rho_{b11}(s)=\rho_{b11}^m(s)+\rho_{b11}^q(s)$ where
\begin{eqnarray}
\rho_{b11}^m(s)&=&\frac{\alpha_sG(1,1)N_c!C_F(D-2)s^{D-2}}{16(4\pi)^D(D-1)^2}
  \int_z^1\left(-\frac{\hat\rho_{a+b}(x)}{x(1-x)}+2\hat L_b(x)\frac{d}{dx}
  \right)\hat\rho_{a1}^m(z/x)dx,\nonumber\\
\rho_{b11}^q(s)&=&\frac{\alpha_sG(1,1)N_c!C_F(D-2)s^{D-2}}{16(4\pi)^D(D-1)^2}
  \int_z^1\left(-\frac{\hat\rho_a(x)}{x(1-x)}+2\hat L_b(x)\frac{d}{dx}\right)
  \hat\rho_{a1}^q(z/x)dx\qquad
\end{eqnarray}
and where again $z=m^2/s$. We have introduced the functions
\begin{equation}
\hat\rho_a(z)=(1+z)\hat\rho_V(1,1;z),\qquad
\hat\rho_b(z)=\left(\frac{D+2}2-\frac{D-2}2z\right)\hat\rho_V(1,1;z),
\end{equation}
$\Big(\hat\rho_{a+b}(z)=\hat\rho_a(z)+\hat\rho_b(z)\Big)$, and
\begin{equation}
\hat L_b(z)=\int_0^z\frac{\hat\rho_b(z')dz'}{(1-z')^2}.
\end{equation}
The details of the calculation can be found in Appendix~B. 

\subsection{The semi-massive fish (c21)}
The two diagrams obtained by connecting the massive line with one of the
massless lines (see \ Fig.~\ref{fig1}(c21))are called semi-massive fish diagrams. Summing up the results
for these two diagrams we obtain
\begin{eqnarray}
\lefteqn{\rho_{c21}(s)\ =\ \frac{g_s^2s^{D/2-3}}{16(4\pi)^{3D/2}(D-2)(D-1)^2}
  \int_{m^2}^sds_1s_1^{D-4}(s-s_1)\hat\rho_V(1,1;s_1/s)\
  \times}\nonumber\\&&\kern-12pt\times\
  \Bigg[4(D-2)\left(D-2+\frac{Ds_1}s\right)
  \left(\hat\rho_m(z_1)\tr_1(\Gamma,s_1)+\hat\rho'_m(z_1)\tr'_1(\Gamma,s_1)
  \right)\,+\nonumber\\&&
  +4(D-2)\left(1-\frac{s_1}s\right)
  \left(\hat\rho_m(z_1)\tr_2(\Gamma,s_1)+\hat\rho'_m(z_1)\tr'_2(\Gamma,s_1)
  \right)\,+\nonumber\\&&
  +\left(D-2+2(D-2)\frac{s_1}s+(D+2)\frac{s_1^2}{s^2}\right)
  \left(\hat\rho_q(z_1)\tr_3(\Gamma,s_1)+\hat\rho'_q(z_1)\tr'_3(\Gamma,s_1)
  \right)\,+\nonumber\\&&
  +\left(1-\frac{s_1}s\right)\left(1+\frac{s_1}s\right)
  \left(\hat\rho_q(z_1)\frac12\left(\tr_4(\Gamma,s_1)+\tr_5(\Gamma,s_1)\right)
  +\hat\rho'_q(z_1)\tr'_4(\Gamma,s_1)\right)\,+\nonumber\\&&
  -\left(1-\frac{s_1}s\right)^2
  \left(\hat\rho_q(z_1)\frac12\left(\tr_4(\Gamma,s_1)+\tr_5(\Gamma,s_1)\right)
  +\hat\rho'_q(z_1)\tr'_5(\Gamma,s_1)\right)\,+\nonumber\\&&
  +\left(1-\frac{s_1}s\right)\left(1+\frac{s_1}s\right)
  \left(\hat\rho_q(z_1)\tr_6(\Gamma,s_1)+\hat\rho'_q(z_1)\tr'_6(\Gamma,s_1)
  \right)\,+\nonumber\\&&
  +4(D-1)\frac{s_1}s\left(\hat\rho''_q(z_1)\frac12\left(\tr_4(\Gamma,s_1)
  +\tr_5(\Gamma,s_1)\right)+\hat\rho'''_q(z_1)\tr'_5(\Gamma,s_1)\right)\Bigg]
\end{eqnarray}
where $z_1=m^2/s_1$ and
\begin{eqnarray}
\hat\rho_m(z_1)&=&2(D-1)B_1,\qquad \hat\rho'_m(z_1)\ =\ B_2,\nonumber\\[7pt]
\hat\rho_q(z_1)&=&2(D-2)\left((D-2+Dz_1)B_4-2(D-1)z_1B_6\right)
  -D(DB_5+(D-2)B_7)\nonumber\\[7pt]
\hat\rho''_q(z_1)&=&2(D-2)\left((1-z_1)B_4+(D-1)(B_8+z_1B_6)\right)
  +D(B_5+(D-2)B_7),\nonumber\\[7pt]
\hat\rho'_q(z_1)&=&DB_5+(D-2)B_7,\qquad\hat\rho'''_q(z_1)\ =\ -B_5
\end{eqnarray}
with
\begin{eqnarray}
\lefteqn{B_1=-\hat\rho_V(1,0,0,1,1;z_1)
  +2(1-z_1)\hat\rho_V(1,0,1,1,1;z_1)\,+}\nonumber\\&&
  +(1+z_1)\hat\rho_V(1,1,1,0,1;z_1)-\hat\rho_V(1,1,1,1,0;z_1)
  +(1-z_1)^2\hat\rho_V(1,1,1,1,1;z_1),\nonumber\\[3pt]
\lefteqn{B_2=-\hat\rho_V(1,0,0,1,1;z_1)+\hat\rho_V(1,0,1,0,1;z_1)
  -\hat\rho_V(1,0,1,1,0;z_1)\,+}\nonumber\\&&
  -\hat\rho_V(1,1,1,-1,1;z_1)+\hat\rho_V(1,1,1,0,0;z_1)
  -(1-z_1)\hat\rho_V(1,1,1,1,0;z_1),\nonumber\\[3pt]
\lefteqn{B_4=-2\hat\rho_V(1,0,0,1,1;z_1)
  +2(1-z_1)\hat\rho_V(1,0,1,1,1;z_1)\,+}\nonumber\\&&
  +2(1+z_1)\hat\rho_V(1,1,1,0,1;z_1)-2\hat\rho_V(1,1,1,1,0;z_1)
  +(1-z_1)^2\hat\rho_V(1,1,1,1,1;z_1),\nonumber\\[3pt]
\lefteqn{B_5=2\hat\rho_V(1,-1,0,1,1;z_1)
  +2\hat\rho_V(1,0,-1,1,1;z_1)\,+}\nonumber\\&&
  -2\hat\rho_V(1,0,0,1,0;z_1)+2(1-z_1)\hat\rho_V(1,0,0,1,1)
  -2(1-z_1)\hat\rho_V(1,0,1,0,1;z_1)\,+\nonumber\\&&
  +2(1-z_1)\hat\rho_V(1,0,1,1,0;z_1)-2z_1\hat\rho_V(1,1,1,-1,1;z_1)
  +2(1+z_1)\hat\rho_V(1,1,1,0,0;z_1)+\nonumber\\&&
  -\hat\rho_V(1,1,1,1,-1;z_1)+(1-z_1)^2\hat\rho_V(1,1,1,1,0;z_1),
  \nonumber\\[3pt]
\lefteqn{B_6=2\hat\rho_V(1,1,1,0,1;z_1)-\hat\rho_V(1,1,1,1,0;z_1),}
  \nonumber\\[3pt]
\lefteqn{B_7=-2\hat\rho_V(1,0,0,1,1)+2\hat\rho_V(1,0,1,0,1;z_1)
  -2\hat\rho_V(1,0,1,1,0;z_1)\,+}\nonumber\\&&
  -2\hat\rho_V(1,1,1,0,0;z_1)+\hat\rho_V(1,1,1,1,-1;z_1)
  -2(1-z_1)\hat\rho_V(1,1,1,1,0;z_1),\nonumber\\[3pt]
\lefteqn{B_8=2\hat\rho_V(1,0,0,1,1;z_1)-2z_1\hat\rho_V(1,1,1,0,1;z_1)
  +\hat\rho_V(1,1,1,1,0;z_1).}
\end{eqnarray}
The spectral functions $\hat\rho_V(n_1,n_2,n_3,n_4,n_5;z_1)$ are so-called
prototypes which are defined in Appendix~C. The additional traces that do not
appear in Eq.~(\ref{traces}) are given by
\begin{eqnarray}\label{tracesp}
\tr'_1(\Gamma,q^2)&=&\frac m{4q^2}\Big\{\Tr(\Gamma\slq[\sigma_{\mu\nu},
  \bar\Gamma]\slq)\sigma^{\mu\nu}-\sigma^{\mu\nu}\Tr([\sigma_{\mu\nu},\Gamma]
  \slq\bar\Gamma\slq)\Big\},\nonumber\\
\tr'_2(\Gamma,q^2)&=&\frac m4\Big\{\Tr(\Gamma\gamma_\alpha
  [\sigma_{\mu\nu},\bar\Gamma]\gamma^\alpha)\sigma^{\mu\nu}-\sigma^{\mu\nu}
  \Tr([\sigma_{\mu\nu},\Gamma]\gamma_\alpha\bar\Gamma\gamma^\alpha)\Big\},
  \nonumber\\
\tr'_3(\Gamma,q^2)&=&\frac1{4q^2}\Big\{\Tr(\Gamma\slq[\sigma_{\mu\nu},
  \bar\Gamma]\slq)\slq\sigma^{\mu\nu}-\sigma^{\mu\nu}\Tr([\sigma_{\mu\nu},
  \Gamma]\slq\bar\Gamma\slq)\slq\Big\},\nonumber\\
\tr'_4(\Gamma,q^2)&=&\frac14\Big\{\Tr(\Gamma\slq\sigma_{\mu\nu}\bar\Gamma
  \gamma_\alpha)\gamma^\alpha\sigma^{\mu\nu}-\gamma^\alpha\Tr(\Gamma
  \gamma_\alpha\bar\Gamma\sigma_{\mu\nu}\slq)\sigma^{\mu\nu}\nonumber\\&&\qquad
  +\sigma^{\mu\nu}\Tr(\Gamma\sigma_{\mu\nu}\slq\bar\Gamma\gamma_\alpha)
  \gamma^\alpha-\sigma^{\mu\nu}\gamma^\alpha\Tr(\Gamma\gamma_\alpha\bar\Gamma
  \slq\sigma_{\mu\nu})\Big\},\nonumber\\
\tr'_5(\Gamma,q^2)&=&\frac14\Big\{\gamma^\alpha\Tr(\Gamma\gamma_\alpha
  \sigma_{\mu\nu}\bar\Gamma\slq)\sigma^{\mu\nu}-\Tr(\Gamma\slq\bar\Gamma
  \sigma_{\mu\nu}\gamma_\alpha)\gamma^\alpha\sigma^{\mu\nu}\nonumber\\&&\qquad
  +\sigma^{\mu\nu}\gamma^\alpha\Tr(\Gamma\sigma_{\mu\nu}\gamma_\alpha
  \bar\Gamma\slq)-\sigma^{\mu\nu}\Tr(\Gamma\slq\bar\Gamma\gamma_\alpha
  \sigma_{\mu\nu})\gamma^\alpha\Big\},\nonumber\\
\tr'_6(\Gamma,q^2)&=&\frac14\Big\{\Tr(\Gamma\gamma_\alpha
  [\sigma_{\mu\nu},\bar\Gamma]\gamma^\alpha)\slq\sigma^{\mu\nu}-\sigma^{\mu\nu}
  \Tr([\sigma_{\mu\nu},\Gamma]\gamma_\alpha\bar\Gamma\gamma^\alpha)\slq\Big\}.
\end{eqnarray}
The traces $\tr'_i(\Gamma,q^2)$ correspond to traces which are absent in the
leading order term. They enter the calculation in the course of renormalizing 
the results. 

\subsection{Renormalization}
The baryonic currents need to be renormalized. In general, there will be 
mixing under renormalization  and therefore one has to construct the
whole matrix of renormalization constants. Within our technique of
parameterizing the results for arbitrary $\Gamma$ matrices this is a
straightforward procedure. The genuine first-order vertex correction for the
current $j$ is given by
\begin{equation}
j^B=\eps^{abc}(u_a^TC(\gamma_\mu\gamma_\nu\Gamma
  +\Gamma\gamma_\nu\gamma_\mu)d_b)\gamma^\mu\gamma^\nu\Gamma'\Psi_c
\end{equation}
which (using $\gamma^\mu\gamma^\nu=g^{\mu\nu}-i\sigma^{\mu\nu}$) can be
written as
\begin{eqnarray}
j^B&=&\eps^{abc}\Big\{(u_a^TC(g_{\mu\nu}\Gamma+\Gamma g_{\mu\nu})d_b)
  \gamma^\mu\gamma^\nu\Gamma'\Psi_c-i(u_a^TC(\sigma_{\mu\nu}\Gamma-\Gamma
  \sigma_{\mu\nu})d_b)\gamma^\mu\gamma^\nu\Gamma'\Psi_c\Big\}\ =\nonumber\\
  &=&\eps^{abc}\Big\{2D(u_a^TC\Gamma d_b)\Gamma'\Psi_c-(u_a^TC(\sigma_{\mu\nu}
  \Gamma-\Gamma\sigma_{\mu\nu})d_b)\sigma^{\mu\nu}\Gamma'\Psi_c\Big\}.
\end{eqnarray}
In calculating the correlator we therefore expect objects of the form
\begin{eqnarray}\label{renobj}
\frac1{4D^2}\langle j_1^Bj_2^B\rangle
  &=&\Tr(\Gamma\circ\bar\Gamma\ \circ)\ \circ\ -\frac1{2D}\Tr(\Gamma\circ
  (\sigma_{\mu\nu}\bar\Gamma-\bar\Gamma\sigma_{\mu\nu})\ \circ)
  \circ\sigma^{\mu\nu}\,+\nonumber\\&&\qquad\qquad
  +\frac1{2D}\sigma^{\mu\nu}\Tr((\sigma_{\mu\nu}\Gamma
  -\Gamma\sigma_{\mu\nu})\circ\bar\Gamma\ \circ)\ \circ\,.
\end{eqnarray}
The open circles stand for further Dirac structures in the calculation. For 
example, if the first term on the r.h.s. of (\ref{renobj}) is given by 
$\Tr(\Gamma\circ\bar\Gamma\ \circ)\ \circ=q^2\tr_3
=\Tr(\Gamma\slq\bar\Gamma\slq)\slq$, the whole right hand side of 
Eq.~(\ref{renobj}) reads
\begin{eqnarray}
\lefteqn{\Tr(\Gamma\slq\bar\Gamma\slq)\slq-\frac1{2D}\Tr\left(\Gamma\slq
  (\sigma_{\mu\nu}\bar\Gamma-\bar\Gamma\sigma_{\mu\nu})\slq\right)\slq
  \sigma^{\mu\nu}}\nonumber\\&&\strut
  +\frac1{2D}\sigma^{\mu\nu}\Tr\left((\sigma_{\mu\nu}\Gamma
  -\Gamma\sigma_{\mu\nu})\slq\bar\Gamma\slq\right)\slq
  \ =\ q^2\left(\tr_3-\frac2D\tr'_3\right).
\end{eqnarray}
where we have used the trace definitions (\ref{traces}) and (\ref{tracesp}). 
The left hand side of Eq.~(\ref{renobj}) represents the singular contribution
of the diagram. On the other hand, the singular parts of the spectral
functions of the basic structures $\tr_i$ are $2\alpha_s/3\pi\eps$ times the
LO term, whereas the spectral functions of the primed structures
$\tr'_i$ are $-\alpha_s/6\pi\eps$ times the LO result of the
corresponding basic structure. Note, finally, that we need to consider only 
the LO $\eps$ singularities within the $\msbar$-scheme. Therefore, we need not
specify $\Gamma$ at this point. If we write the total result for the
semi-massive fish as
\begin{equation}
\sum_{i=1}^6\left[\rho_b^i\tr_i+\rho_b^{i\prime}\tr'_i\right]
  =\sum_{i=1}^6\left[\left(\rho_{00}^i+\rho_{01}^i\eps+\frac{\alpha_s}\pi
  \Big(\rho_{10}^i\frac1\eps+\rho_{11}^i\Big)\right)\tr_i
  +\frac{\alpha_s}\pi\Big(\rho_{10}^{i\prime}\frac1\eps+\rho_{11}^{i\prime}
  \Big)\tr'_i\right]
\end{equation}
with $\rho_{10}^i=\frac23\rho_{00}^i$ and
$\rho_{10}^{i\prime}=-\frac16\rho_{00}^i$, we can extract the renormalization
factors and obtain
\begin{equation}
\sum_{i=1}^6\left[\left(1+\frac{2\alpha_s}{3\pi\eps}\right)
  \left(\rho_{00}^i+\rho_{01}^i\eps+\frac{\alpha_s}\pi\Big(\rho_{11}^i
  -\frac23\rho_{01}^i\Big)\right)\tr_i+\left(-\frac{\alpha_s}{6\pi\eps}
  \rho_{00}^{i}+\frac{\alpha_s}\pi\rho_{11}^{i\prime}\right)\tr'_i
  \right].
\end{equation}
Up to $O(\alpha_s)$ we have
\begin{equation}
\rho_r^i=\rho_{00}^i+\rho_{01}^i\eps+\frac{\alpha_s}\pi
  \Big(\rho_{11}^i-\frac23\rho_{01}^i\Big).
\end{equation}
If we substitute this expression for $\rho_{00}^i$ in the coefficient of
$\tr'_i$, the next-to-leading order contribution can be skipped while the term
proportional to $\eps$ leads to a subtraction of the finite term,
\begin{equation}
\sum_{i=1}^6\left[\left(1+\frac{2\alpha_s}{3\pi\eps}\right)
  \rho_r^i\tr_i+\left(-\frac{\alpha_s}{6\pi\eps}\rho_r^i
  +\frac{\alpha_s}\pi\Big(\rho_{11}^{i\prime}+\frac16\rho_{01}^i\right)\tr'_i
  \right]\,.
\end{equation}
In total we have
\begin{equation}
\rho_b^i\tr_i+\rho_b^{i\prime}\tr'_i
  =\left(1+\frac{2\alpha_s}{3\pi\eps}\right)\rho_r^i\tr_i
  +\left(-\frac{\alpha_s}{6\pi\eps}\rho_r^i+\rho_r^{i\prime}\right)\tr'_i
\end{equation}
where
\begin{equation}
\rho_r^{i\prime}=\frac{\alpha_s}\pi\Big(\rho_{11}^{i\prime}
  +\frac16\rho_{01}^i\Big).
\end{equation}
The coefficient of the primed structure has no LO contribution.
Nevertheless, the renormalization works in the same formal manner as for the
basic structure. As these calculations show, the renormalization of the mixing
operators for the basic and primed structure is accomplished by a
{\em renormalization matrix\/},
\begin{equation}
\pmatrix{\rho_b^i\cr\noalign{\smallskip}\rho_b^{i\prime}\cr}\ =\
  \pmatrix{\displaystyle 1+\frac{2\alpha_s}{3\pi\eps}&0\cr\noalign{\smallskip}
  \displaystyle-\frac{\alpha_s}{6\pi\eps}&1\cr}
  \pmatrix{\rho_r^i\cr\noalign{\smallskip}\rho_r^{i\prime}\cr}.
\end{equation}
One can easily invert this renormalization matrix to compute the renormalized 
quantities. One also needs the bare quanties which are given by
\begin{equation}
\rho_r^i=\left(1-\frac{2\alpha_s}{3\pi\eps}\right)\rho_b^i,\qquad
\rho_r^{i\prime}=\rho_b^{i\prime}+\frac{\alpha_s}{6\pi\eps}\rho_b^i.
\end{equation}
The results given in Appendices~A to C are already renormalized ones. In order
to obtain the total result we have to combine the spectral functions giving
\begin{equation}
\hat\rho^i_0(z)=\hat\rho_{\rm leading}^i(z),\qquad
\hat\rho^i_1(z)=\hat\rho_{\rm light}^i(z)
  +\hat\rho_{\rm massi}^i(z)+\hat\rho_{\rm fish}^i(z),\qquad
\hat\rho^{i\prime}_1(z)=\hat\rho_{\rm fish}^{i\prime}(z).
\end{equation}
Using these spectral functions and the traces defined in Eqs.~(\ref{traces})
and~(\ref{tracesp}), the spectral density is given by
\begin{eqnarray}\label{rhofull}
\rho(s)&=&\frac{s^2}{512\pi^4}\sum_{i=1}^6
  \Bigg[\hat\rho^i_0(m^2/s)\left\{1+\left[\left(n_m^i+
  \frac{3-4r_\Gamma+r_\Gamma^2}3\right)\frac{\alpha_s}\pi\tr_i
  -\frac{\alpha_s}{6\pi}\tr'_i\right]\ln\pfrac{m^2}{\mu^2}\right\}
  +\strut\nonumber\\&&\qquad\qquad\qquad\qquad\strut
  +\frac{\alpha_s}\pi\left(\hat\rho^i_1(m^2/s)\tr_i
  +\hat\rho^{i\prime}_1(m^2/s)\tr'_i\right)\Bigg]
\end{eqnarray}
where $n_m^i=0,1$ depending on whether there is a factor of $m$ in $\tr_i$ 
or not. Explicitly one has $n_m^{1,2}=1$ and $n_m^{3-6}=0$.

\section{Some properties of the spectral densities at NLO}
In this section we consider two limiting cases of the NLO spectral densities.
First we consider the large energy or equivalently the mass zero limit. 
Second we analyze the near threshold limit relevant for a comparison with 
HQET results. Both limits are interesting and physically relevant. The
limiting cases for the lowest spin baryons have been discussed before in
Refs.~\cite{masspart,mompart}. We therefore concentrate on the case of the
``vector'' current~(\ref{veccur}) in the following.

\subsection{High energy expansion}
In the high energy (or, equivalently, small mass) limit $z\rightarrow 0$ the
spectral density reads
\begin{eqnarray}
\lefteqn{\rho^{\mu\nu}(s)=\frac{s^2}{512\pi^4}\Bigg[
  4\left\{1+\frac{\alpha_s}\pi\left(\frac{59}{12}+\ln\pfrac{\mu^2}s
  \right)\right\}m\frac{q^\mu q^\nu}s
  -3\left\{1+\frac{\alpha_s}\pi\left(\frac{13}3+\frac23\ln\pfrac{\mu^2}s
  \right)\right\}mg^{\mu\nu}+\strut}\nonumber\\&&\strut
  +\frac45\left\{1+\frac{\alpha_s}\pi\left(\frac{343}{180}-\frac13
  \ln\pfrac{\mu^2}s\right)\right\}\slq\frac{q^\mu q^\nu}s
  +\frac15\left\{1+\frac{\alpha_s}\pi\left(\frac{521}{60}+3\ln\pfrac{\mu^2}s
  \right)\right\}q^\mu\gamma^\nu+\strut\nonumber\\&&\strut
  +\frac15\left\{1+\frac{\alpha_s}\pi\left(\frac{313}{180}-\frac13
  \ln\pfrac{\mu^2}s\right)\right\}\gamma^\mu q^\nu
  -\frac45\left\{1+\frac{\alpha_s}\pi\left(\frac{82}{45}-\frac13
  \ln\pfrac{\mu^2}s\right)\right\}\slq g^{\mu\nu}+\strut\nonumber\\&&\strut
  -\frac{\alpha_s}\pi\left(\frac{25}{36}+\frac13\ln\pfrac{\mu^2}s\right)
  \left(2m\slq\frac{\gamma^\mu q^\nu-q^\mu\gamma^\nu}s
  +\slq\gamma^\mu\gamma^\nu\right)-\frac{\alpha_s}\pi\left(\frac94
  +\ln\pfrac{\mu^2}s\right)m\gamma^\mu\gamma^\nu\Bigg]\qquad
\label{ziszero}
\end{eqnarray}
where $m=m_{\msbar}(\mu)$ is the $\msbar$ mass. When one compares
(\ref{ziszero}) with the results of an {\it ab initio} (multiplicatively 
renormalized) massless calculation one does not obtain full agreement. 
We discuss an alternative route.
Instead of expanding $\rho^{\mu\nu}$ along the set of ten covariants given in 
Eq.~(\ref{vectordecomp}) we exploit the symmetry of the problem and
expand along an alternative set of 
only nine covariants
\[\Big\{mq^\mu q^\nu/s,mg^{\mu\nu},\slq q^\mu q^\nu/s,
  \gamma^\nu q^\mu,\gamma^\mu q^\nu,\slq g^{\mu\nu},
  m\slq\gamma^\mu\gamma^\nu\slq/s,m\gamma^\mu\gamma^\nu,
  \gamma^\mu\slq\gamma^\nu\Big\}.\]
In this case we obtain
\begin{eqnarray}
\lefteqn{\rho^{\mu\nu}(s)=\frac{s^2}{512\pi^4}\Bigg[
  4\left\{1+\frac{\alpha_s}\pi\left(\frac{59}{12}+\ln\pfrac{\mu^2}s\right)
  \right\}m\frac{q^\mu q^\nu}s
  -3\left\{1+\frac{\alpha_s}\pi\left(\frac{13}3+\frac23\ln\pfrac{\mu^2}s
  \right)\right\}mg^{\mu\nu}+\strut}\nonumber\\&&\strut
  +\frac45\left\{1+\frac{\alpha_s}\pi\left(\frac{343}{180}-\frac13
  \ln\pfrac{\mu^2}s\right)\right\}\slq\frac{q^\mu q^\nu}s
  +\frac15\left\{1+\frac{\alpha_s}\pi\left(\frac{313}{180}-\frac13
  \ln\pfrac{\mu^2}s\right)\right\}q^\mu\gamma^\nu+\strut\nonumber\\&&\strut
  +\frac15\left\{1+\frac{\alpha_s}\pi\left(\frac{313}{180}-\frac13
  \ln\pfrac{\mu^2}s\right)\right\}\gamma^\mu q^\nu
  -\frac45\left\{1+\frac{\alpha_s}\pi\left(\frac{82}{45}-\frac13
  \ln\pfrac{\mu^2}s\right)\right\}\slq g^{\mu\nu}+\strut\nonumber\\&&\strut
  -\frac{\alpha_s}\pi\left(\frac{25}{36}+\frac13\ln\pfrac{\mu^2}s\right)
  \left(m\frac{\slq\gamma^\mu\gamma^\nu\slq}s-\gamma^\mu\slq\gamma^\nu\right)
  -\frac{\alpha_s}\pi\left(\frac{14}9+\frac23\ln\pfrac{\mu^2}s\right)
  m\gamma^\mu\gamma^\nu\Bigg]\qquad
\end{eqnarray}
where the first six contributions which contain both LO and NLO parts are
reproduced by the massless calculation. It cannot be expected that a
multiplicatively renormalized massless result can reproduce the remaining
three contributions. In addition, the massless calculation cannot reproduce
terms such as $z\ln(z)$ as they appear for instance in the full NLO results in
Eqs.~(\ref{corr1m}) and~(\ref{corr1q}). These terms can be parametrized with
condensates of local operators\footnote{For a discussion of this point cf.\
Ref.~\cite{masspart}.}.

\subsection{Near-threshold expansion}
As concerns the $\Lambda$-type baryons we have already compared the 
near-threshold limit with the HQET result in Ref.~\cite{masspart}. Here we 
repeat the exercise for the $\Sigma$--type $J^{P}=1/2^{+}$ baryon with 
$\Gamma=\gamma^\mu$, $\Gamma'=\gamma_\mu\gamma_5$, i.e. $s_\Gamma=-1$, and 
$r_\Gamma=1$. In the 
near-threshold limit $E\rightarrow 0$ ($s=(m+E)^2$) Eq.~(\ref{rhofull}) gives
\begin{eqnarray}\label{hqet0}
\rho^{\mu\nu}_{\rm thr}(m,E)&=&\frac{E^5}{40\pi^4m}\Bigg[
  2\left\{1+\frac{\alpha_s}\pi\left(\frac{42}5+\frac{4\pi^2}9
  +2\ln\pfrac\mu{2E}\right)\right\}(m+\slq)\frac{q^\mu q^\nu}s
  +\strut\nonumber\\&&\strut
  -\left\{1+\frac{\alpha_s}\pi\left(\frac{121}{15}+\frac{4\pi^2}9
  +2\ln\pfrac\mu{2E}+\frac13\ln\pfrac{m^2}{\mu^2}\right)\right\}
  (m+\slq)g^{\mu\nu}+\strut\nonumber\\&&\strut
  -\frac{\alpha_s}{3\pi}(q^\mu\gamma^\nu+\gamma^\mu q^\nu)
  -\frac{\alpha_s}{6\pi}\left(1-\ln\pfrac{m^2}{\mu^2}\right)
  m\frac{\slq\gamma^\mu\gamma^\nu\slq}s+\strut\nonumber\\&&\strut
  -\frac{\alpha_s}{6\pi}\left(1-\ln\pfrac{m^2}{\mu^2}\right)
  m\gamma^\mu\gamma^\nu
  +\frac{\alpha_s}{3\pi}\left(1-\ln\pfrac{m^2}{\mu^2}\right)
  \gamma^\mu\slq\gamma^\nu\Bigg]
\end{eqnarray}
where $m$ is the pole mass. The HQET contribution is obtained by projecting on
both sides with $(1+\slv)/2=(m+\slq)/2m$. Omitting the projector $(1+\slv)/2$
itself we obtain
\begin{eqnarray}\label{hqet1}
\rho^{\mu\nu}_{\rm HQET}(m,E)&=&\frac{E^5}{20\pi^4m}\Bigg[
  2\left\{1+\frac{\alpha_s}\pi\left(\frac{42}5+\frac{4\pi^2}9
  +2\ln\pfrac\mu{2E}\right)\right\}\frac{q^\mu q^\nu}s
  +\strut\nonumber\\&&\strut
  -\left\{1+\frac{\alpha_s}\pi\left(\frac{121}{15}+\frac{4\pi^2}9
  +2\ln\pfrac\mu{2E}+\frac13\ln\pfrac{m^2}{\mu^2}\right)\right\}g^{\mu\nu}
  +\strut\nonumber\\&&\strut
  -\frac{\alpha_s}{3\pi}\frac{q^\mu q^\nu}s
  -\frac{\alpha_s}{6\pi}\left(1-\ln\pfrac{m^2}{\mu^2}\right)
  \left(\gamma^\mu\gamma^\nu+\frac{\gamma^\mu q^\nu-q^\mu\gamma^\nu}m\right)
  +\strut\nonumber\\&&\strut
  +\frac{\alpha_s}{6\pi}\left(1-\ln\pfrac{m^2}{\mu^2}\right)
  \left(\frac{q^\mu q^\nu}s-\gamma^\mu\gamma^\nu+\frac{q^\mu\gamma^\nu
  -\gamma^\mu q^\nu}m\right)\Bigg].
\end{eqnarray}
We now proceed to compare (\ref{hqet1}) with the corresponding result derived 
from the HQET current~\cite{Grozin:1992td}
\begin{equation}
\tilde j_{\Sigma1}=(q^TC\gamma_{\perp}^{\mu} q)\gamma^{\perp}_{\mu}
\gamma_5\tilde Q\,,
\end{equation}
where $\gamma_{\perp}^{\mu}=\gamma_{\nu}(g^{\mu\nu}-q^\mu q^\nu/q^2)$. One
needs to extract the transverse piece from Eq.~(\ref{hqet1}) using the 
transverse projection $g_{\mu\nu}-q_\mu q_\nu/q^2$.
\begin{equation}\label{hqet2}
m\rho_{\Sigma1}(m,E)=-\frac{3E^5}{20\pi^4}
  \left\{1+\frac{\alpha_s}\pi\left(\frac{136}{15}+\frac{4\pi^2}9
  +2\ln\pfrac\mu{2E}-\frac23\ln\pfrac{m^2}{\mu^2}\right)\right\}.
\end{equation}
This can be compared with the HQET result~\cite{Groote:1996em}
\begin{equation}\label{hqet3}
\tilde\rho_{\Sigma1}(E,\mu)=-\frac{3E^5}{20\pi^4}
  \left\{1+\frac{\alpha_s}\pi\left(\frac{116}{15}+\frac{4\pi^2}9
  +2\ln\pfrac\mu{2E}\right)\right\}
\end{equation}
where the explicit mass factor $m$ appearing on the left hand side of
Eq.~(\ref{hqet2}) has been absorbed in the definition of
$\tilde\rho_{\Sigma1}(E,\mu)$ in Eq.~(\ref{hqet3}).\footnote{Note that we have
to take into account the contraction with $\bar\Gamma'\cdots\Gamma'$ as well.}
After addition of the matching coefficient~\cite{Grozin:1992td}
\begin{equation}\label{matchcoef}
C_{\Sigma1}(m/\mu,\alpha_s)
  =1+\frac{\alpha_s}{3\pi}\left(2-\ln\pfrac{m^2}{\mu^2}\right)
\end{equation}
one obtains $m\rho_{\Sigma1}(m,E)=C_{\Sigma1}(m/\mu,\alpha_s)^2
  \tilde\rho_{\Sigma1}(\mu,E)$.
In this case the matching procedure allows one to recover the near-threshold
limit of the full correlator starting from the simpler effective theory near
threshold~\cite{Eichten:1989zv}. Note that the higher order $E/m$ corrections 
to Eq.~(\ref{hqet0}) can be easily obtained from the explicit result
given in Eq.~(\ref{corr1q}). Indeed, the next-to-leading order correction in
the low energy threshold expansion reads
\begin{eqnarray}
\Delta\rho_{\Sigma1}(m,E)&=&-\frac{11E^6}{120m^2}\left\{1+\frac{\alpha_s}\pi
  \left(\frac{1744}{165}+\frac{4\pi^2}9-\frac53\ln\pfrac{m^2}{\mu^2}
  +\frac{74}{33}\ln\pfrac m{2E}\right)\right\}m+\strut\nonumber\\&&\strut
  -\frac{17E^6}{120m^2}\left\{1+\frac{\alpha_s}\pi\left(\frac{512}{51}
  +\frac{4\pi^2}9-\frac{28}{51}\ln\pfrac{m^2}{\mu^2}
  +\frac{110}{51}\ln\pfrac m{2E}\right)\right\}\slq.
\end{eqnarray}
To obtain this result starting from HQET is a more difficult task requiring
the analysis of contributions from higher dimension operators. 

\subsection{Interpolation}

\begin{figure}[htb]\begin{center}
\epsfig{figure=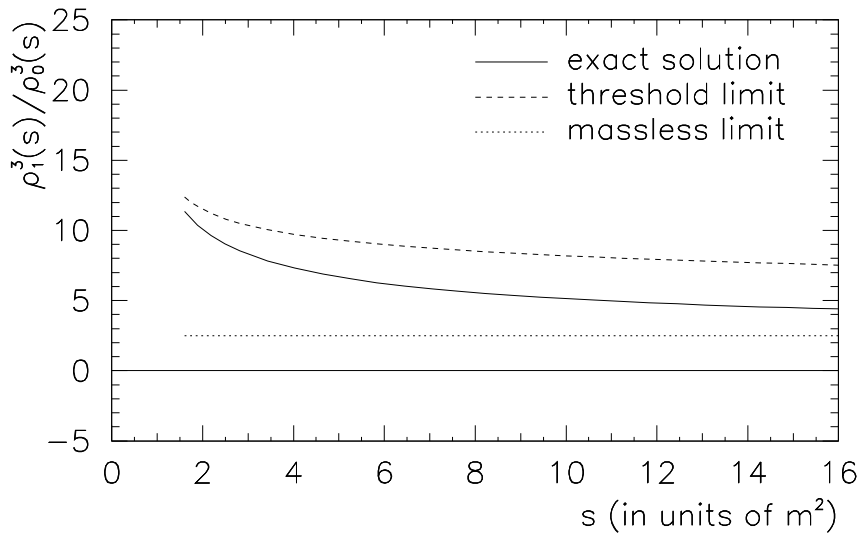, scale=0.7}
\caption{\label{ratio3}The ratio $\rho^3_1(s)/\rho^3_0(s)$ up to NLO
as a function of the squared energy $s$.}
\end{center}\end{figure}

We now discuss some quantitative features of the correction given in
Eq.~(\ref{rhofull}). Of interest is whether the two limiting expressions
(the massless and threshold limit) can be used
to characterize the full function for all energies. To this end we
compare components of the baryonic spectral density up to
NLO. In Fig.~\ref{ratio3} we show the ratio
$\rho^i_1(s)/\rho^i_0(s)$ for $i=3$. Fig.~\ref{ratio3} shows that one would
obtain a rather good approximation for the full NLO order
correction for all values of $s$ if one were to use an interpolation between
the two limiting cases.

\subsection{Moments}

An instructive set of observables are the negative moments of the spectral 
density (see e.g.~\cite{Groote:2001im})\footnote{We have normalized $M^{i}_n$
such that $M^{i}_n=1$ for $\hat\rho^{\,i}(z)=1$.},
\begin{eqnarray}\label{defmom}
{\cal M}_{-n}&=&\int_{m^2}^\infty s^{-n}\rho(s)ds
  =\frac{m^{6-2n}}{512\pi^4(n-3)}\sum_{i=1}^6\left(M_{n-4}^i\tr_i
  +M_{n-4}^{i\prime}\tr'_i\right),\nonumber\\
M^{i(\prime)}_n&=&(n+1)\int_0^1z^n\hat\rho^{i(\prime)}(z)dz
\end{eqnarray}
where $\rho(s)$ is taken from Eq.~(\ref{rhofull}) and $\rho^i(m^2/s)$ and
$\rho^{i\prime}(m^2/s)$ are the coefficients of $\tr_i$ and $\tr'_i$. One has
\begin{eqnarray}\label{resform1}
M_n^i&=&M_n^{i(0)}\left\{1+\frac{\alpha_s}\pi\left(\left(n_m^i
  +\frac{3-4r_\Gamma+r_\Gamma^2}3\right)\ln\pfrac{\mu^2}{m^2}
  +\delta_n^i\right)\right\},\nonumber\\
M_n^{i\prime}&=&M_n^{i(0)}\left\{-\frac{\alpha_s}{6\pi}\ln\pfrac{\mu^2}{m^2}
  +\frac{\alpha_s}\pi\delta_n^{i\prime}\right\},
\end{eqnarray}
where 
\begin{eqnarray}
M_n^{1(0)}&=&\frac{12}{(n+2)^2(n+3)(n+4)(n+5)},\nonumber\\
M_n^{2(0)}&=&\frac6{(n+2)^2(n+3)^2(n+4)(n+5)},\nonumber\\
M_n^{3(0)}&=&\frac{12}{(n+2)(n+3)(n+4)(n+5)(n+6)},\nonumber\\
M_n^{4(0)}&=&\frac6{(n+2)(n+3)^2(n+4)(n+5)(n+6)}
  \ =\ M_n^{5(0)}\ =\ M_n^{6(0)}\qquad\quad
\end{eqnarray}
and $\delta_n^i=\delta_n^{i(a)}+\delta_n^{i(b)}+\delta_n^{i(c)}$ with
\begin{eqnarray}
\delta_n^{1(a)}&=&\frac23(2-4r_\Gamma+r_\Gamma^2)\left(\psi(n+1)+\gamma_E
  -\frac{3n+1}{4(n+1)(n+2)}\right)+\frac23r_\Gamma\ =\ \delta_n^{2(a)},
  \nonumber\\
\delta_n^{3(a)}&=&\frac23(2-4r_\Gamma+r_\Gamma^2)\left(\psi(n+1)+\gamma_E
  -\frac{n(3n+5)}{4(n+1)(n+2)}\right)+\frac23r_\Gamma
  \ =\ \delta_n^{4(a)}=\delta_n^{5(a)}=\delta_n^{6(a)},\nonumber\\
\delta_n^{1(b)}&=&\frac43\left(\psi(n+1)+\gamma_E+\frac{n^2-2}{2(n+1)(n+2)}
  \right)\ =\ \delta_n^{2(b)},\nonumber\\
\delta_n^{3(b)}&=&\frac43\left(\psi(n+3)+\gamma_E+\frac{2n^2+2n+1}{4(n+1)(n+2)}
  \right)\ =\ \delta_n^{4(b)}=\delta_n^{5(b)}=\delta_n^{6(b)},
  \nonumber\\
\delta_n^{1(c)}&=&\frac43\Bigg(\frac{n^3+3n^2-n-5}{(n+1)(n+2)(n+3)}
  (\psi(n+1)+\gamma_E)+\strut\nonumber\\&&\qquad\strut
  +\frac{3n^3+8n^2-3n-12}{2(n+1)^2(n+2)(n+3)}-\psi'(n+2)\Bigg)+\frac{4\pi^2}9
  \ =\ \delta_n^{2(c)},\nonumber\\
\delta_n^{3(c)}&=&\frac43\Bigg(\frac{n^5+13n^4+55n^3+75n^2-26n-82}{(n+1)(n+2)
  (n+3)(n+4)(n+5)}(\psi(n+1)+\gamma_E)+\strut\nonumber\\&&\strut
  +\frac{n^6+16n^5+122n^4+448n^3+655n^2+120n-300}{4(n+1)^2(n+2)(n+3)(n+4)(n+5)}
  -\psi'(n+2)\Bigg)+\frac{4\pi^2}9\ =\ \delta_n^{6(c)},\nonumber\\
\delta_n^{4(c)}&=&\frac43\Bigg(\frac{-2(n^3+9n^2+24n+19)}{(n+1)(n+2)(n+3)(n+4)}
  (\psi(n+1)+\gamma_E)+\strut\nonumber\\&&\qquad\strut
  -\frac{n^6+8n^5+16n^4-12n^3-21n^2+108n+132}{4(n+1)^2(n+2)(n+3)(n+4)}
  -\psi'(n+2)\Bigg)+\frac{4\pi^2}9\ =\ \delta_n^{5(c)},\nonumber\\
\delta_n^{1\prime}&=&-\frac2{3(n+4)}\left(\psi(n+1)+\gamma_E
  +\frac{n^2-2n-2}{4(n+1)}\right)\ =\ \delta_n^{2\prime},\nonumber\\
\delta_n^{3\prime}&=&-\frac2{3(n+5)}\left(\psi(n+1)+\gamma_E
  +\frac{n^2-n-5}{4(n+2)}\right)
  \ =\ \delta_n^{4\prime}=\delta_n^{5\prime}=\delta_n^{6\prime}\,.
\end{eqnarray}
Further $\psi(z)=\Gamma'(z)/\Gamma(z)$ is the digamma function, $\psi'(z)$ its
first derivative (first polygamma function) and $\gamma_E=0.577\ldots$ is the
Euler--Mascheroni constant. In order to combine the three terms 
$\delta_n^{i(a)}$, $\delta_n^{i(b)}$ and $\delta_n^{i(c)}$ into a closed form 
expression we use the recursion relations
\begin{equation}
\psi(n+1)=\psi(n)+\frac1n,\quad\psi(1)=-\gamma_E,\qquad
\psi'(n+1)=\psi'(n)-\frac1{n^2},\quad\psi'(1)=\frac{\pi^2}6\, .
\end{equation}
Note that the three terms $\delta_n^{i(a)}$, $\delta_n^{i(b)}$ and
$\delta_n^{i(c)}$ stem from different diagrams. With the help of the results 
from Appendices~A to~C, resp., one obtains
\begin{equation}
\label{delta}
\delta_n^i=A_n^i+\frac{2\pi^2}9,\qquad\delta_n^{i\prime}=A_n^{i\prime}\,.
\end{equation}
The coefficients $A_n^i$ and $A_n^{i\prime}$ are rational numbers. As an
example we list the first values for $A_n^3$ in the first parts of the three
columns of Table~\ref{tab1} for $r_\Gamma=0$ (scalar current), $r_\Gamma=1$
(vector current), and for the sake of completeness $r_\Gamma=2$ (tensor
current, $\Gamma=\sigma^{\mu\nu}$).
\begin{table}\begin{center}
\begin{tabular}{|l||r|r||r|r||r|r|}\hline
  $n$&\multicolumn{2}{|c||}{$r_\Gamma=0$ (scalar)}&
  \multicolumn{2}{|c||}{$r_\Gamma=1$ (vector)}&
  \multicolumn{2}{|c|}{$r_\Gamma=2$ (tensor)}\\
  &$A_n^3$&$\delta_n^3-\delta_{n-1}^3$
  &$A_n^3$&$\delta_n^3-\delta_{n-1}^3$
  &$A_n^3$&$\delta_n^3-\delta_{n-1}^3$\\\hline
  $0$&$2.66667$&$-$&$3.33333$&$-$&$4.00000$&$-$\\
  $1$&$5.59259$&$2.92593$&$4.92593$&$1.59259$&$5.14815$&$1.14815$\\
  $2$&$7.39815$&$1.80556$&$5.98148$&$1.05556$&$5.95370$&$0.80556$\\
  $3$&$8.70741$&$1.30926$&$6.75741$&$0.77593$&$6.55185$&$0.59815$\\
  $4$&$9.73056$&$1.02315$&$7.36389$&$0.60648$&$7.01944$&$0.46759$\\
  $5$&$10.5674$&$0.83683$&$7.85786$&$0.49397$&$7.39913$&$0.37968$\\
  $6$&$11.2735$&$0.70611$&$8.27230$&$0.41444$&$7.71635$&$0.31722$\\
  $7$&$11.8831$&$0.60957$&$8.62791$&$0.35561$&$7.98730$&$0.27095$\\
  $8$&$12.4186$&$0.53552$&$8.93842$&$0.31052$&$8.22281$&$0.23552$\\
  $9$&$12.8956$&$0.47701$&$9.21341$&$0.27499$&$8.43046$&$0.20765$\\
\hline\end{tabular}
\caption{\label{tab1}Values for the rational part $A_n^3$ of the first moments
  $\delta_n^3$ and their relative difference $\delta_n^3-\delta_{n-1}^3$ for
  the scalar, vector and tensor current}
\end{center}\end{table}
We represent the moments in the form
\begin{equation}\label{resform1norm}
\frac{M_n^i}{M_n^{i(0)}}=\frac{M_N^i}{M_N^{i(0)}}
\left\{1+\frac{\alpha_s}\pi(\delta_n^i-\delta_N^i)\right\},\qquad
\frac{M_n^{i\prime}}{M_n^{i(0)}}=\frac{M_N^{i\prime}}{M_N^{i(0)}}
  \left\{\frac{\alpha_s}\pi(\delta_n^{i\prime}-\delta_N^{i\prime})\right\}.
\end{equation}
As mentioned before, all moments defined in (\ref{defmom}) are normalized to 
one in the sense that $M^{i}_n=1$ for $\hat\rho^{\,i}(z)=1$. Note that the 
difference $\delta_n-\delta_N$
is scheme-independent. This feature was used in the high precision analysis of
heavy quark properties~\cite{bbmass} within NRQCD (see e.g.\
Ref.~\cite{Hoang:2000yr}). One can now easily find the actual magnitude of
the correction. Indeed, for any desired precision and range of $n$, a set of
perturbatively commensurate moments can be found. We therefore present
differences of consecutive $\delta_n^i$ in the second part of the columns of
Table~\ref{tab1}.

Note that because of the $s$-integration in Eq.~(\ref{defmom}), moments
represent massive vacuum bubbles, i.e.\ diagrams without external momenta with
massive lines. These diagrams have been comprehensively analyzed in
Refs.~\cite{Avdeev:1995eu,Broadhurst:1998rz}. The analytical results for the
first few moments at the three-loop level can be checked independently with
existing computer programs (see e.g.\ Ref.~\cite{Chetyrkin:1997mb}).

\section{Applications to physics: Sum rules}
A phenomenological application is given by QCD sum rules (see e.g.\
Ref.~\cite{Ioffe:kw,Chung:wm}) or finite energy sum rules (FESR, see e.g.\
Ref.~\cite{zphys}). In general, the main quantity of interest is the residue
\begin{equation}\label{residuedef}
\langle|j_B(0)|\Lambda(p,\sigma)\rangle=\lambda_B u(p,\sigma)
\end{equation}
where $\Lambda(p,\sigma)\rangle$ is a baryon state and $u(p,\sigma)$ is the
spinor that satisfies the free equation of motion $(\slp-m_B)u(p,\sigma)=0$.
The contribution of the ground state baryon $B$ to the spectral density of the
correlator reads
\begin{equation}
\rho_{B(0)}(s)=\lambda^2\delta(s-m_B^2)
\end{equation}
while the excited states contribute to the spectral density $\rho_B(s)$ that
were calculated before. We define a threshold value $s_0$ where the excited
states and the continuum start contributing. If we take moments, i.e.\ 
integrals over the phenomenological spectral density
$\rho_{B(0)}(s)+\theta(s-s_0)\rho_B(s)$ with different
powers of $s$, we obtain the sum rule condition
\begin{equation}
\int_0^\infty\rho_B(s)s^nds=\int_0^\infty\left(\rho_{B(0)}(s)
  +\theta(s-s_0)\rho_B(s)\right)s^nds=\lambda_B^2m_B^{2n}
  +\int_{s_0}^\infty\rho_B(s)s^nds
\end{equation}
or, equivalently, the sum rule
\begin{equation}
\lambda_B^2m_B^{2n}=\int_0^{s_0}\rho_B(s)s^nds=:{\cal M}_n(s_0).
\end{equation}
The two unknown parameters $s_0$ and $\lambda_B$ ($m_B$ is assumed to be known)
can be determined in turn by the sum rule analysis. For this purpose we
calculate the ratio of nearby moments,
\begin{equation}
\frac{{\cal M}_n(s_0)}{{\cal M}_{n-1}(s_0)}=\frac{\int_0^{s_0}\rho_B(s)s^nds}
  {\int_0^{s_0}\rho_B(s)s^{n-1}ds}=m_B^2
\end{equation}
and adjust $s_0$. Once $s_0$ is determined we can calculate $\lambda_B^2$
by using the zeroth moment,
\begin{equation}
\lambda_B^2={\cal M}_0(s_0)=\int_0^{s_0}\rho_B(s)ds.
\end{equation}
The second mass scale occuring in the problem is the mass $m$ of the heavy
constituent quark occuring in $\rho_B(s)$. It is fixed by assuming a value of
$500\MeV$ for the difference $\Lambda_B=m_B-m$ between baryon and quark mass.
Finally, we have to decide which of the sum rules we use. Here we have decided
to use the $n=0$ sum rule. For $\rho(s)$ we use Eq.~(\ref{rhofull}) in the
scalar setting ($\Lambda$-type baryon: $\Gamma=1$, $\Gamma'=\gamma_5$) or
vector settings ($\Sigma$-type baryon: $\Gamma=\gamma^\mu$,
$\Gamma'=\gamma_\mu\gamma_5$, $\Sigma^*$-type baryon: $\Gamma=\gamma^\mu$,
$\Gamma'=1$ with $\Psi\to\Psi^\mu$). $\alpha_s$ is the running QCD coupling
with a scale set by the heavy quark mass.

\subsection{$\Lambda$-type baryons}
The $\Lambda$-type baryon ground state masses are given by~\cite{PDG}
\begin{equation}
m_{\Lambda_c^+}=2286.46\pm 0.14\MeV,\qquad
m_{\Lambda_b^0}=5624\pm 9\MeV.
\end{equation}
We first do the sum rule analysis for the LO contribution. Using the sum rule
\begin{equation}
\frac{\displaystyle\int_{m^2}^{s_0}\rho^{\rm LO}(s)ds}{\displaystyle
  \int_{m^2}^{s_0}\rho^{\rm LO}(s)\frac{ds}s}
  =m^2\frac{\displaystyle\int_{z_0}^1\hat\rho_0(z)\frac{dz}{z^4}}{\displaystyle
  \int_{z_0}^1\hat\rho_0(z)\frac{dz}{z^3}}=m_\Lambda^2
\end{equation}
with $z_0=m^2/s_0$, we obtain the sum rule value for $z_0$. Reinserting this
value into ${\cal M}_0(s_0)$, we obtain the value for 
$\lambda_{\Lambda_c^+}^2$. The results are collected in Tab.~\ref{sumlambda}.
\begin{table}[ht]\begin{center}
\begin{tabular}{|r||l|l||l|l|}\hline
  $\Lambda_c^+$&$z_0$ (LO)&$z_0$ (NLO)&$\lambda$ (LO)&$\lambda$ (NLO)\\\hline
$m$-part&$0.565550$&$0.564193$
  &$1.77978\times 10^{-3}\GeV^3$&$3.01893\times 10^{-3}\GeV^3$\\
$q$-part&$0.564191$&$0.562752$
  &$1.60523\times 10^{-3}\GeV^3$&$2.72316\times 10^{-3}\GeV^3$\\\hline
\end{tabular}\vspace{12pt}
\begin{tabular}{|r||l|l||l|l|}\hline
  $\Lambda_b^0$&$z_0$ (LO)&$z_0$ (NLO)&$\lambda$ (LO)&$\lambda$ (NLO)\\\hline
$m$-part&$0.805403$&$0.804903$
  &$1.81465\times 10^{-3}\GeV^3$&$2.86564\times 10^{-3}\GeV^3$\\
$q$-part&$0.805088$&$0.804575$
  &$1.74339\times 10^{-3}\GeV^3$&$2.75323\times 10^{-3}\GeV^3$\\\hline
\end{tabular}
\end{center}
\caption{\label{sumlambda}Sum rule analysis for the $\Lambda$-type baryons
  with $E_\Lambda=m_\Lambda-m=500\MeV$}
\end{table}
In going from LO to NLO, the value of $z_0$ does not change significantly.
However, the values for $\lambda_{\Lambda_c^+}$ and $\lambda_{\Lambda_b^0}$ 
change by a factor of $1.7$ ($\Lambda_c^+$)
and $1.6$ ($\Lambda_b^0$) which disqualifies the sum rule method as an
appropriate method to determine the baryon's parameters. Theoretically, we can
improve the convergence behaviour by tuning the mass logarithm $\ln(m^2/\mu^2)$
to values of roughly $4.5$ ($m$-part) or $6.0$ ($q$-part). However, the
necessary values of $\mu$ (less than $200\MeV$) are far from being realistic.

\subsection{$\Sigma$-type baryons}
For the $\Sigma_c$-type baryons, the three ground state masses read~\cite{PDG}
\begin{equation}
m_{\Sigma_c^{++}}=2454.02\pm 0.18\MeV,\quad
m_{\Sigma_c^+}=2452.9\pm 0.4\MeV,\quad
m_{\Sigma_c^0}=2453.76\pm 0.18\MeV.
\end{equation}
We take the mass average
$m_{\Sigma_c}=2.453.56\MeV$ as input parameter. For the recently found
$\Sigma_b$ state we take the average of the mass of the positive and negative
charged state and obtain $m_{\Sigma_b}=5811.5\MeV$~\cite{Aaltonen:2007rw}.
Tab.~\ref{sumsigma} gives an overview over the results. The NLO/LO ratio for
the residue is $1.6$ for $\Sigma_c$ and $1.46$ for $\Sigma_b$.
\begin{table}[ht]\begin{center}
\begin{tabular}{|r||l|l||l|l|}\hline
$\Sigma_c$&$z_0$ (LO)&$z_0$ (NLO)&$\lambda$ (LO)&$\lambda$ (NLO)\\\hline
$m$-part&$0.590168$&$0.589363$
  &$2.99916\times 10^{-3}\GeV^3$&$4.73288\times 10^{-3}\GeV^3$\\
$q$-part&$0.588942$&$0.588086$
  &$2.72806\times 10^{-3}\GeV^3$&$4.30472\times 10^{-3}\GeV^3$\\\hline
\end{tabular}\vspace{12pt}
\begin{tabular}{|r||l|l||l|l|}\hline
$\Sigma_b$&$z_0$ (LO)&$z_0$ (NLO)&$\lambda$ (LO)&$\lambda$ (NLO)\\\hline
$m$-part&$0.811194$&$0.810902$
  &$3.10502\times 10^{-3}\GeV^3$&$4.53653\times 10^{-3}\GeV^3$\\
$q$-part&$0.810897$&$0.810598$
  &$2.98752\times 10^{-3}\GeV^3$&$4.36492\times 10^{-3}\GeV^3$\\\hline
\end{tabular}\vspace{12pt}
\caption{\label{sumsigma}Sum rule analysis for the $\Sigma$-type baryons
  with $E_{\Sigma}=m_\Sigma-m=500\MeV$}
\end{center}\end{table}

\subsection{$\Sigma^*$-type baryons}

In order to extract the vector baryon components of highest spin $3/2$ we use
the Rarita--Schwinger formalism (see e.g.~\cite{Kaloshin:2003xc}). Instead of
the spinor $u(p,\sigma)$ in Eq.~(\ref{residuedef}) we have to use the
Rarita--Schwinger spinor$u_\mu(p,\sigma)$ ($\sigma$ takes the four values
$\pm 3/2$ and $\pm 1/2$). The equation of motion for the highest spin state
leads to the two additional constraints
\begin{equation}
\gamma^\mu u_\mu(p,\sigma)=0\qquad\mbox{and}\qquad
p^\mu u_\mu(p,\sigma)=0.
\end{equation}
Using the ansatz
\begin{eqnarray}
\Pi^{\mu\nu}=Aq^\mu q^\nu+Bg^{\mu\nu}+\frac1i\sigma^{\alpha\beta}
  D_{\alpha\beta}^{\mu\nu}
\end{eqnarray}
where
\begin{equation}
D_{\alpha\beta}^{\mu\nu}=R(g_\alpha^\mu g_\beta^\nu-g_\alpha^\nu g_\beta^\mu)
  +X_1q^\mu(g_\alpha^\nu q_\beta-g_\beta^\nu q_\alpha)
  +X_2q^\nu(g_\alpha^\mu q_\beta-g_\beta^\mu q_\alpha),
\end{equation}
we can contract from the left with $\gamma^\mu$ in order to satisfy the first
constraint $\gamma_\mu\Pi^{\mu\nu}=0$. We obtain $X_1=-X_2$, $A+4X_2=0$ and
\begin{equation}
R=\frac16\left(\frac A2q^2-B\right).
\end{equation}
Further imposing the condition $q_\mu\Pi^{\mu\nu}$ ~\cite{Kirchbach:2004qg},
we can determine all coefficients up to one for which we choose $A$. This 
gives rise to a
basis of projectors $({\cal P}^{3/2})^{\mu\nu}$,
$({\cal P}^{1/2}_{11})^{\mu\nu}$, and $({\cal P}^{1/2}_{22})^{\mu\nu}$ and
nilpotent operators $({\cal P}^{1/2}_{12})^{\mu\nu}$ and
$({\cal P}^{1/2}_{21})^{\mu\nu}$ of the
form~\cite{Kaloshin:2003xc,Pilling:2004cu}
\begin{eqnarray}
({\cal P}^{3/2})^{\mu\nu}
  &=&g^{\mu\nu}-\frac23\frac{q^\mu q^\nu}{q^2}-\frac13\gamma^\mu\gamma^\nu
  +\frac1{3q^2}(\gamma^\mu q^\nu-\gamma^\nu q^\mu)\slq,\nonumber\\
({\cal P}^{1/2}_{11})^{\mu\nu}
  &=&\frac13\gamma^\mu\gamma^\nu-\frac13\frac{q^\mu q^\nu}{q^2}
  -\frac1{3q^2}(\gamma^\mu q^\nu-\gamma^\nu q^\mu)\slq,\nonumber\\
({\cal P}^{1/2}_{22})^{\mu\nu}&=&\frac{q^\mu q^\nu}{q^2},\nonumber\\
({\cal P}^{1/2}_{12})^{\mu\nu}&=&\sqrt{\frac3{q^2}}\
  \frac1{3q^2}(-q^\mu+\gamma^\mu\slq)\slq q^\nu,\nonumber\\
({\cal P}^{1/2}_{21})^{\mu\nu}&=&\sqrt{\frac3{q^2}}\
  \frac1{3q^2}q^\mu(-q^\nu+\gamma^\nu\slq)\slq.
\end{eqnarray}
In order to obtain the highest spin component we contract with
$({\cal P}^{3/2})^{\mu\nu}$. If we use the above results for a sum rule
analysis of the $\Sigma^*$-type baryons with ground state masses~\cite{PDG}
\begin{equation}
m_{\Sigma_c^{*++}}=2518.4\pm 0.6\MeV,\quad
m_{\Sigma_c^{*+}}=2517.5\pm 2.3\MeV,\quad
m_{\Sigma_c^{*0}}=2518.0\pm 0.5\MeV
\end{equation}
with an average value of $m_{\Sigma_c}=2518.0\MeV$, we obtain the values in
Table~\ref{sumsigmas}. For the recently found $\Sigma_b^*$-baryon we take the 
average of the mass of the positive and negative charged state and obtain
$m_{\Sigma_b^{*}}=5832.7\MeV$ \cite{Aaltonen:2007rw}). The ratio NLO/LO is
$1.55$ for $\Sigma_c^*$ and $1.45$ for $\Sigma_b^*$.
\begin{table}[ht]\begin{center}
\begin{tabular}{|r||l|l||l|l|}\hline
$\Sigma_c^*$&$z_0$ (LO)&$z_0$ (NLO)&$\lambda$ (LO)&$\lambda$ (NLO)\\\hline
$m$-part&$0.599077$&$0.598274$
  &$1.22636\times 10^{-3}\GeV^3$&$1.90952\times 10^{-3}\GeV^3$\\
$q$-part&$0.597895$&$0.597045$
  &$1.11838\times 10^{-3}\GeV^3$&$1.74130\times 10^{-3}\GeV^3$\\\hline
\end{tabular}\vspace{12pt}
\begin{tabular}{|r||l|l||l|l|}\hline
$\Sigma_b^*$&$z_0$ (LO)&$z_0$ (NLO)&$\lambda$ (LO)&$\lambda$ (NLO)\\\hline
$m$-part&$0.811837$&$0.811537$
  &$1.26773\times 10^{-3}\GeV^3$&$1.83958\times 10^{-3}\GeV^3$\\
$q$-part&$0.811543$&$0.811234$
  &$1.21994\times 10^{-3}\GeV^3$&$1.77024\times 10^{-3}\GeV^3$\\\hline
\end{tabular}
\caption{\label{sumsigmas}Sum rule analysis for the $\Sigma^*$-type baryons
  with $E_{\Sigma^*}=m_\Sigma^*-m=500\MeV$}
\end{center}\end{table}

\section{Conclusions}
To conclude, we have computed the NLO perturbative corrections to the 
correlators of finite mass baryons containing one heavy quark and two massless
quarks for a variety of quantum numbers of the baryonic currents. Technically 
this is a genuine three loop calculation with the two mass 
scales $s$ and $m^{2}$. We have considered both the massless limit and the 
threshold HQET limit of the correlator as special cases of the general finite 
mass formula. The two respective limiting expressions agree with previous
massles and HQET results in the literature. From threshold to high energies 
the exact spectral density interpolates nicely between the leading order
(leading in $1/m_{Q}$) HQET 
result close to threshold and the asymptotic mass zero result. This raises 
the hope that one can find a similar interpolation formula at the four loop 
NNLO level using the massless and the HQET four-loop results. These can be 
calculated using existing computational 
algorithms~\cite{ibyparts,Chetyrkin:xa}.

\subsection*{Acknowledgements}
This work was supported in part by the Estonian target financed project
No.~0182647s04, by the Estonian Science Foundation under grant No.~6216, by
the RFFI grant No.~06-02-16659 and by the DFG grant under contract No.\
DFG~SI~349/10-1. S.~Groote and A.A.~Pivovarov acknowledge partial support from
the Deutsche Forschungsgemeinschaft (436EST 17/1/06 and 436RUS 17/68/06).
\begin{appendix}

\section{The light contributions}
\setcounter{equation}{0}\def\theequation{A\arabic{equation}}
The light contribution consists of contributions from the two diagrams
(b21, light self energy) and (c11, light fish). These contributions are
calculated in turn.

\subsection{The light self energy (b21)}
The result has the same structure as the leading order contribution,
\begin{equation}
\rho_{b21}(s)=\frac{g_s^2G^2}{16(4\pi)^{3D/2}}s^{3D/2-4}
  \sum_{i=1}^6\hat\rho_{b21}^i(m^2/s)\tr_i(\Gamma,s)
\end{equation}
where $\tr_i(\Gamma,s)$ are given by Eqs.~(\ref{traces}) and
\begin{eqnarray}
\hat\rho_{b21}^1(z)&=&\frac{D-2}{2(D-1)^2}\left(\frac1{\eps^2}
  -\frac1{6\eps}+\frac{17}{12}\right)\hat\rho_{a1}^{1*}(z),\nonumber\\
\hat\rho_{b21}^2(z)&=&\frac{D-2}{2(D-1)^2}\left(\frac1{\eps^2}
  -\frac1{6\eps}+\frac{17}{12}\right)\hat\rho_{a1}^{2*}(z)
  -\frac23\left(\frac1\eps+\frac52\right)\hat\rho_{a1}^{m*}(z),
  \nonumber\\
\hat\rho_{b21}^3(z)&=&\frac{D-2}{2(D-1)^2}\left(\frac1{\eps^2}
  -\frac1{6\eps}+\frac{17}{12}\right)\hat\rho_{a1}^{3*}(z),\nonumber\\
\hat\rho_{b21}^4(z)&=&\frac{D-2}{2(D-1)^2}\left(\frac1{\eps^2}
  -\frac1{6\eps}+\frac{17}{12}\right)\hat\rho_{a1}^{4*}(z),\nonumber\\
\hat\rho_{b21}^5(z)&=&\frac{D-2}{2(D-1)^2}\left(\frac1{\eps^2}
  -\frac1{6\eps}+\frac{17}{12}\right)\hat\rho_{a1}^{5*}(z),\nonumber\\
\hat\rho_{b21}^6(z)&=&\frac{D-2}{2(D-1)^2}\left(\frac1{\eps^2}
  -\frac1{6\eps}+\frac{17}{12}\right)\hat\rho_{a1}^{6*}(z)
  -\frac23\left(\frac1\eps+\frac52\right)\hat\rho_{a1}^{q*}(z)
\end{eqnarray}

\subsection{The light fish (c11)}
For the light fish the traces are more complicated. However, using the
parameter $c_\Gamma$ introduced in Eq.(\ref{cgamma}), the result again has
again the same structure as the leading order contribution,
\begin{equation}
\rho_{c11}(s)=\frac{g_s^2G^2}{16(4\pi)^{3D/2}}s^{3D/2-4}
  \sum_{i=1}^6\hat\rho_{c11}^i(m^2/s)\tr_i(\Gamma,s)
\end{equation}
where
\begin{eqnarray}
\hat\rho_{c11}^1(z)&=&\left(\frac6\eps+19-24\zeta(3)
  +\frac{c_\Gamma^2}{12}\left(\frac6{\eps^2}+\frac2\eps-1\right)\right)
  \rho_{a1}^{1*}(z),\nonumber\\
\hat\rho_{c11}^2(z)&=&\left(\frac6\eps+19-24\zeta(3)
  +\frac{c_\Gamma^2}{12}\left(\frac6{\eps^2}+\frac2\eps-1\right)\right)
  \rho_{a1}^{2*}(z)-\left(2+c_\Gamma^2\left(\frac1{3\eps}+1\right)\right)
  \rho_{a1}^{m*}(z),\nonumber\\
\hat\rho_{c11}^3(z)&=&\left(\frac6\eps+19-24\zeta(3)
  +\frac{c_\Gamma^2}{12}\left(\frac6{\eps^2}+\frac2\eps-1\right)\right)
  \rho_{a1}^{3*}(z),\nonumber\\
\hat\rho_{c11}^4(z)&=&\left(\frac6\eps+19-24\zeta(3)
  +\frac{c_\Gamma^2}{12}\left(\frac6{\eps^2}+\frac2\eps-1\right)\right)
  \rho_{a1}^{4*}(z),\nonumber\\
\hat\rho_{c11}^5(z)&=&\left(\frac6\eps+19-24\zeta(3)
  +\frac{c_\Gamma^2}{12}\left(\frac6{\eps^2}+\frac2\eps-1\right)\right)
  \rho_{a1}^{5*}(z),\nonumber\\
\hat\rho_{c11}^6(z)&=&\left(\frac6\eps+19-24\zeta(3)
  +\frac{c_\Gamma^2}{12}\left(\frac6{\eps^2}+\frac2\eps-1\right)\right)
  \rho_{a1}^{6*}(z)-\left(2+c_\Gamma^2\left(\frac1{3\eps}+1\right)\right)
  \rho_{a1}^{q*}(z).\qquad\quad
\end{eqnarray}
Note that the parameter $c_\Gamma$ only appears in squared form, so that
the signature of the current (i.e.\ whether or not there is a $\gamma_5$)
is irrelevant for the result.

\subsection{Merger with the leading order diagram}
When one combines the results for the two light diagrams and adds the leading
order contribution, the divergences can be factored out. When one combines the
contributions one has to take into account colour and combinatorial factors.
The relevant factors are $N_c!$ for the leading order diagram (a1),
$2N_c!C_F$ for the light self energy diagram (b21), and $-N_c!C_B$ for the
light fish (c11). The results read
\begin{eqnarray}
\rho_{\rm leading}(s)&=&N_c!\rho_{a1}(s)
  \ =\ N_c!\frac{(D-2)Gs^{2-2\eps}}{16(4\pi)^D(D-1)^2}\
  \frac1\eps\hat\rho_{a1}(m^2/s),\nonumber\\
\rho_{\rm light}(s)&=&\frac{\alpha_sN_c!C_F}{4\pi}
  \left\{\left(\frac{B_0}{\eps^2}+\frac{B_1}\eps+B_2\right)\rho^*_{a1}(s)
  +\left(\frac{B'_1}\eps+B'_2\right)\rho^{\prime *}_{a1}(s)\right\}
  \ =\nonumber\\
  &=&\frac{\alpha_sN_c!C_F}{4\pi}\Bigg\{
  \frac{(D-2)G^2s^{2-3\eps}}{16(4\pi)^D(D-1)^2}
  \left(\frac{B_0}{\eps^2}+\frac{B_1}\eps+B_2\right)\hat\rho^*_{a1}(m^2/s)
  \,+\nonumber\\&&\qquad\qquad\qquad
  +\frac{G^2s^{2-3\eps}}{2(4\pi)^DD}\left(\frac{B'_1}\eps+B'_2\right)
  \hat\rho^{\prime *}_{a1}(m^2/s)\Bigg\}
\end{eqnarray}
where the coefficients $B_i$, $B'_i$ are given in Eq.~(\ref{bcoeffs}), and
\begin{eqnarray}
\hat\rho_{a1}(m^2/s)&:=&\sum_{i=1}^6\hat\rho_{a1}^i(m^2/s)\tr_i(\Gamma,s),
  \nonumber\\
\hat\rho^*_{a1}(m^2/s)&:=&\sum_{i=1}^6\hat\rho_{a1}^{i*}(m^2/s)\tr_i(\Gamma,s),
  \nonumber\\[7pt]
\hat\rho_{a1}^{\prime*}(m^2/s)&:=&\hat\rho_{a1}^{m*}(m^2/s)\tr_2(\Gamma,s)
  +\hat\rho_{a1}^{q*}(m^2/s)\tr_6(\Gamma,s).
\end{eqnarray}
The functions $\hat\rho_{a1}(z)$, $\hat\rho^*_{a1}(z)$, and
$\hat\rho_{a1}^{\prime *}(z)$ that occur in the spectral densities for the
leading and light contributions vanish for $\eps\rightarrow 0$. This can be
seen by retracing the construction down to the elements
$\hat\rho_{a1}^i(z)$ and $\hat\rho_{a1}^{i*}(z)$ ($i=1,2,\ldots,6$). These
elements are given by a linear combination of spectral functions
$\hat\rho_V(1,n\eps-p;z)$ as defined in Eq.~(\ref{rhovdef}). Using the
abbreviations
\begin{equation}
C_n\eps:=\frac1{\Gamma(n\eps-1)\Gamma(3-(n+1)\eps)},\qquad
C_1^p:=\frac{\Gamma(n\eps-1)\Gamma(3-(n+1)\eps)}{\Gamma(n\eps-p)
  \Gamma(p+2-(n+1)\eps)}
\end{equation}
for the overall and relative factor, the elements can now be written as
\begin{eqnarray}
\hat\rho_{a1}^i(z)=C_1\eps\hat g_1^i(z),&&
\hat\rho_{a1}^{i*}(z)=C_2\eps\hat g_2^i(z),\nonumber\\[3pt]
\hat\rho_{a1}^{\prime i}(z)=C_1\eps\hat g_1^{\prime i}(z),&&
\hat\rho_{a1}^{\prime i*}(z)=C_2\eps\hat g_2^{\prime i}(z),\qquad
i=1,2,\ldots,6
\end{eqnarray}
where (note that $C_n^1=1$)
\begin{eqnarray}
\hat g_n^1(z)&=&D\Bigg\{C_n^2\hat g_{n2}(z)+C_n^0(1-z)^2\hat g_{n0}(z)
  -2\left(\frac{D-2}D-z\right)\hat g_{n1}(z)\Bigg\},\nonumber\\
\hat g_n^2(z)&=&-\Bigg\{C_n^2\hat g_{n2}(z)+C_n^0(1-z)^2\hat g_{n0}(z)
  +2\left(\frac{3D-4}{D-2}+z\right)\hat g_{n1}(z)\Bigg\},\nonumber\\
\hat g_n^3(z)&=&\frac{D+2}2\Bigg\{C_n^3\hat g_{n3}(z)
  +C_n^0\left(\frac{D-2}{D+2}+z\right)(1-z)^2\hat g_{n0}(z)
  \,+\nonumber\\&&\qquad\qquad
  +C_n^2\left(\frac{6-D}{D+2}+3z\right)\hat g_{n2}(z)
  +\left(\frac{2-D}{D+2}-\frac{2Dz}{D+2}+3z^2\right)\hat g_{n1}(z)\Bigg\},
  \nonumber\\
\hat g_n^4(z)&=&-\frac12\Bigg\{C_n^3\hat g_{n3}(z)
  -C_n^0(1-z)^3\hat g_{n0}(z)
  +C_n^2(1+3z)\hat g_{n2}(z)
  -(1-z)(1+3z)\hat g_{n1}(z)\Bigg\},\nonumber\\
\hat g_n^5(z)&=&-\frac12\Bigg\{C_n^3\hat g_{n3}(z)
  -C_n^0(1-z)^3\hat g_{n0}(z)
  +C_n^2(1+3z)\hat g_{n2}(z)
  -(1-z)(1+3z)\hat g_{n1}(z)\Bigg\},\nonumber\\
\hat g_n^6(z)&=&-\frac12\Bigg\{C_n^3\hat g_{n3}(z)
  +C_n^0(1-z)^2(1+z)\hat g_{n0}(z)\,+\nonumber\\&&\qquad
  +C_n^2\left(\frac{7D-10}{D-2}+3z\right)\hat g_{n2}(z)
  +\left(\frac{7D-10}{D-2}+2\frac{3D-4}{D-2}z+3z^2\right)
  \hat g_{n1}(z)\Bigg\}.\nonumber\\[3pt]
\hat g_n^{\prime 1}(z)&=&\hat g_n^{\prime 3}(z)\ =\ \hat g_n^{\prime 4}(z)
  \ =\ \hat g_n^{\prime 5}(z)\ =\ 0,\nonumber\\[3pt]
\hat g_n^{\prime 2}(z)&=&-\hat g_{n1}(z),\nonumber\\
\hat g_n^{\prime 6}(z)&=&-\frac12\left\{(1+z)\hat g_{n1}(z)
  +C_n^2\hat g_{n2}(z)\right\}.
\end{eqnarray}
We can cast this into a more compact form by writing
\begin{equation}
\hat\rho_{a1}^{(\prime)}(z)=C_1\eps\hat g_1^{(\prime)}(z),\qquad
\hat\rho_{a1}^{(\prime)*}(z)=C_2\eps\hat g_2^{(\prime)}(z)
\end{equation}
where
\begin{equation}
\hat g_n^{(\prime)}(m^2/s)=\sum_{i=1}^6\hat g_n^{(\prime)i}(m^2/s)
  \tr_i(\Gamma,s).
\end{equation}
These expressions can then be inserted into the sum of leading and light
contributions. One obtains
\begin{eqnarray}
\rho_l(s)&:=&\rho_{\rm leading}(s)+\rho_{\rm light}(s)
  \ =\ \frac{GN_c!}{16(4\pi)^D}s^{2-2\eps}\frac{(D-2)}{(D-1)^2}C_1\
  \times\nonumber\\&&\times\ \Bigg\{\hat g_1(m^2/s)
  +\frac{\alpha_sC_F}{4\pi}s^{-\eps}\frac{C_2G}{C_1\eps}(B_0+B_1\eps+B_2\eps^2)
  \hat g_2(m^2/s)\,+\nonumber\\&&\qquad\qquad
  +\frac{\alpha_sC_F}{4\pi}s^{-\eps}\frac{8(D-1)^2C_2G}{D(D-2)C_1}
  (B'_1+B'_2\eps)\hat g_2^\prime(m^2/s)\Bigg\}\ =\\
  &=&\frac{GN_c!}{16(4\pi)^D}s^{2-2\eps}\frac{(D-2)}{(D-1)^2}C_1
  \Bigg\{\hat g_1(m^2/s)+\frac{\alpha_sC_F}{4\pi\eps}C^r\hat g_2(m^2/s)
  +\frac{\alpha_sC_F}{4\pi}C^{r\prime}\hat g_2^\prime(m^2/s)\Bigg\}.\nonumber
\end{eqnarray}
The coefficients $C^r$ and $C^{r\prime}$ can be expanded in $\eps$, viz.
\begin{eqnarray}
C^r&:=&s^{-\eps}\frac{C_2G}{C_1}(B_0+B_1\eps+B_2\eps^2)
  \ =\ C^{r(0)}+C^{r(1)}\eps+O(\eps^2),\nonumber\\
C^{r\prime}&=&s^{-\eps}\frac{8(D-1)^2C_2G}{D(D-2)C_1}(B'_1+B'_2\eps)
  \ =\ C^{r\prime(0)}+O(\eps).
\end{eqnarray}
The main singularity of the NLO contribution is proportional to
$\hat g_2(z)$. But $\hat g_2(z)$ is similar to $\hat g_1(z)$. Indeed, one
obtains
\begin{equation}
\hat g_{21}(z):=\hat g_2(z)-\hat g_1(z)=\hat g_{21}^{(1)}(z)\eps+O(\eps^2).
\end{equation}
Therefore, the leading singular NLO part is added and subtracted where
$\hat g_2(z)$ is replaced by $\hat g_1(z)$. The leading singular part of this
new contribution is then combined with the LO part. Expanding
\begin{equation}
g_2(z)=g_2^{(0)}(z)+O(\eps),\qquad
g_2^\prime(z)=g_2^{\prime(0)}(z)+O(\eps)
\end{equation}
and finally using $D=4$, one obtains
\begin{eqnarray}
\rho_l(s)&=&\frac{GN_c!}{16(4\pi)^D}s^{2-2\eps}\frac{(D-2)}{(D-1)^2}C_1
  \Bigg\{\left(1+\frac{\alpha_sC_F}{4\pi\eps}C^{r(0)}\right)
  \hat g_1(m^2/s)\,+\nonumber\\&&\qquad
  +\frac{\alpha_sC_F}{4\pi}\left(C^{r(1)}\hat g_1(m^2/s)
  +C^{r(0)}\hat g_{21}^{(1)}(m^2/s)+C^{r\prime(0)}\hat g_2^\prime(m^2/s)
  \right)\Bigg\}\ =\nonumber\\
  &=&\frac{2N_c!s^2}{9(4\pi)^4}\Bigg\{\left(1+\frac{\alpha_sC_F}{4\pi\eps}
  C^{r(0)}\right)\hat g_1(m^2/s)\,+\nonumber\\&&\qquad
  +\frac{\alpha_sC_F}{4\pi}\left(C^{r(1)}\hat g_1^{(0)}(m^2/s)+C^{r(0)}
  \hat g_{21}^{(1)}(m^2/s)+C^{r\prime(0)}\hat g_2^{\prime(0)}(m^2/s)\right)
  \Bigg\}.
\end{eqnarray}
The LO contributions $\hat\rho_{a1}^i(z)$ are given by
Eqs.~(\ref{LOfullres:terms}). For the remaining light contributions one has
\begin{equation}
\hat\rho_{\rm light}^i(z)=\frac13(2-4r_\Gamma+r_\Gamma^2)\hat\rho_a^i(z)
  +\frac23r_\Gamma\hat\rho_{a1}^{i0}(z)\,,
\end{equation}
where
\begin{eqnarray}
\hat\rho_a^1(z)&=&\frac{19}{24}+\frac{41}{18}z-\frac{29}6z^2
  +\frac{13}6z^3-\frac{29}{72}z^4-\left(1+\frac{10}3z-6z^2+2z^3-\frac13z^4
  \right)\ln(1-z)+\strut\nonumber\\&&\strut
  +\left(\frac12+\frac73z-6z^2+2z^3-\frac13z^4\right)\ln z
  +4z\left(\Li_2(z)-\Li_2(1)+\frac12\ln^2z\right),\nonumber\\
\hat\rho_a^2(z)&=&\frac{25}{96}+\frac{41}9z-\frac{47}{12}z^2-z^3
  +\frac{29}{288}z^4-\left(\frac14+\frac{11}3z-3z^2-z^3+\frac1{12}z^4\right)
  \ln(1-z)+\strut\nonumber\\&&\strut\kern-5pt
  +\left(\frac18+\frac{19}6z-\frac32z^2-z^3+\frac1{12}z^4\right)\ln z
  +(2z+3z^2)\left(\Li_2(z)-\Li_2(1)+\frac12\ln^2z\right),\nonumber\\
\hat\rho_a^3(z)&=&\frac{107}{600}-\frac{59}{120}z+\frac{47}{30}z^2
  -\frac{67}{30}z^3+\frac{149}{120}z^4-\frac{157}{600}z^5
  +\strut\nonumber\\&&\strut
  -\left(\frac15-z+2z^2-2z^3+z^4-\frac15z^5\right)\ln(1-z)
  +\strut\nonumber\\&&\strut
  +\left(\frac1{10}-\frac12z+2z^2-2z^3+z^4-\frac15z^5\right)\ln z,\nonumber\\
\hat\rho_a^4(z)&=&\hat\rho_a^5(z)
  \ =\ \hat\rho_a^6(z)\ =\ \frac{137}{2400}-\frac{107}{240}z
  -\frac{47}{180}z^2+\frac9{10}z^3-\frac{47}{160}z^4+\frac{157}{3600}z^5
  +\strut\nonumber\\&&\strut
  -\left(\frac1{20}-\frac12z-\frac13z^2+z^3-\frac14z^4+\frac1{30}z^5\right)
  \ln(1-z)+\strut\nonumber\\&&\strut
  +\left(\frac1{40}-\frac14z-\frac13z^2+z^3-\frac14z^4+\frac1{30}z^5\right)
  \ln z-z^2\left(\Li_2(z)-\Li_2(1)+\frac12\ln^2z\right).\nonumber\\
\end{eqnarray}

\section{The self energy correction of the massive line (b11)}
\setcounter{equation}{0}\def\theequation{B\arabic{equation}}
The first order self energy correction of the massive line is given by the
inclusion of the {\em master bubble}
\begin{equation}
\Pi_B(k^2)=\int\dDl\frac1{(l^2-m^2)(k-l)^2}
  =\frac{(m^2)^{D/2-2}}{(4\pi)^{D/2}}V(1,1;k^2/m^2)\, .
\end{equation}
The corresponding spectral density reads
\begin{equation}
\rho_B(s)=\frac{(m^2)^{D/2-2}}{(4\pi)^{D/2}}\rho_V(1,1;s/m^2)
  =\frac{s^{D/2-2}}{(4\pi)^{D/2}}\hat\rho_V(1,1;m^2/s).
\end{equation}
With the help of the dispersive representation it is easy to see that the 
mass and momentum part of the self energy correction can be written as
\begin{eqnarray}
\Sigma_m(k^2)&=&Dg_s^2\int\frac{\rho_B(s)ds}{s-k^2},\nonumber\\
\Sigma_p(k^2)&=&\frac{2-D}2g_s^2\int\left(1+\frac{m^2}s\right)
  \frac{\rho_B(s)ds}{s-k^2}.
\end{eqnarray}
In order to enact the renormalization and to absorb the singular parts of
these NLO contributions in the renormalization factors for the mass and
the wave function, we use an expansion for the propagator--type factor,
\begin{equation}
\frac{i(1+a)}{\slk-m(1+b)}\approx\frac{i}{\slk-m}
  +\frac{i}{\slk-m}\Big(-i\slk a-im(b-a)\Big)\frac{i}{\slk-m}\,.
\end{equation}
One obtains
\begin{eqnarray}
a(k^2)&=&\Sigma_p(k^2)\ =\ \frac{2-D}2g_s^2\int\frac{\rho_B(s)ds}{s-k^2}
  \left(1+\frac{m^2}s\right)\ =:\ \int\frac{\rho_a(s)ds}{s-k^2},\nonumber\\
b(k^2)&=&\Sigma_p(k^2)+\Sigma_m(k^2)
  \ =\ g_s^2\int\frac{\rho_B(s)ds}{s-k^2}\left(\frac{D+2}2
  -\frac{D-2}2\frac{m^2}s\right)\ =:\ \frac{\rho_b(s)ds}{s-k^2}.\qquad
\end{eqnarray}
Using momentum subtraction at $k^2=m^2$, the singular parts can be split off,
\begin{eqnarray}
a(k^2)&=&\int\frac{\rho_a(s)ds}{s-k^2}
  \ =\ \int\frac{\rho_a(s)ds}{s-m^2}+\int\left(\frac1{s-k^2}
  -\frac1{s^2-m^2}\right)\rho_a(s)ds\ =\nonumber\\
  &=&\int\frac{\rho_a(s)ds}{s-m^2}+(k^2-m^2)\int\frac{\rho_a(s)ds}{(s-m^2)
  (s-k^2)}\ =:\ a(m^2)+a_f(k^2),\nonumber\\
b(k^2)&=&\int\frac{\rho_b(s)ds}{s-m^2}+(k^2-m^2)\int\frac{\rho_b(s)ds}{(s-m^2)
  (s-k^2)}\ =:\ b(m^2)+b_f(k^2)\qquad
\end{eqnarray}
where
\begin{eqnarray}
a(m^2)&=&\int\frac{\rho_a(s)ds}{s-m^2}\ =\ \frac{-g_s^2}{(4\pi)^{D/2}}
  \pfrac{\mu^2}{m^2}^\eps\frac G\eps\left(1+\zeta(2)\eps^2+O(\eps^3)\right),
  \nonumber\\
b(m^2)&=&\int\frac{\rho_b(s)ds}{s-m^2}\ =\ \frac{-g_s^2}{(4\pi)^{D/2}}
  \pfrac{\mu^2}{m^2}^\eps\frac G\eps\left(3-2\eps+3\zeta(2)\eps^2+O(\eps^3)
  \right).\qquad
\end{eqnarray}
Having absorbed the divergent parts into the renormalization of mass and wave
function, the finite parts can be expanded again,
\begin{eqnarray}
\frac{i(1+a_f)}{\slk-m(1+b_f)}
  &\approx&\frac{i\slk}{k^2-m^2}\left(1+a_f+\frac{2m^2b_f}{k^2-m^2}\right)
  +\frac{im}{k^2-m^2}\left(1+a_f+b_f+\frac{2m^2b_f}{k^2-m^2}\right)
  \ =\nonumber\\
  &=:&\frac{i\slk}{k^2-m^2}\left(1+P(k^2)\right)
  +\frac{im}{k^2-m^2}\left(1+M(k^2)\right).
\end{eqnarray}
For the leading order diagram one obtains
\begin{equation}
V_{a1}(q^2)=-\frac{2G(1,1)}{(4\pi)^{D/2}}\int\dDk\frac{i}{\slk-m}
  \left(-(q-k)^2\right)^{D/2-1}
\end{equation}
where $G(1,1)=G/\eps$ is the massless master bubble. The corrections to the
propagator results in the use of an effective propagator
\begin{equation}
D_{\rm eff}(k^2)=\frac{i\slk}{k^2-m^2}D_{\rm eff}^k(k^2)
  +\frac{im}{k^2-m^2}D_{\rm eff}^m(k^2)
\end{equation}
with \Big($\rho_{a+b}(s)=\rho_a(s)+\rho_b(s)$\Big)
\begin{eqnarray}
D_{\rm eff}^k(k^2)&=&\frac{i}{k^2-m^2}
  -i\int\frac{\rho_a(s)ds}{(s-m^2)(k^2-s)}
  +2im^2\int\frac{\rho_b(s)ds}{(s-m^2)^2}
  \left(\frac1{k^2-m^2}-\frac1{k^2-s}\right),\nonumber\\
D_{\rm eff}^m(k^2)&=&\frac{i}{k^2-m^2}
  -i\int\frac{\rho_{a+b}(s)ds}{(s-m^2)(k^2-s)}
  +2im^2\int\frac{\rho_b(s)ds}{(s-m^2)^2}
  \left(\frac1{k^2-m^2}-\frac1{k^2-s}\right)\qquad\quad
\end{eqnarray}
One finally obtains
\begin{equation}
V_{a1}(q^2)+V_{b11}(q^2)=-\frac{2G(1,1)}{(4\pi)^{D/2}}\int\dDk
  \frac{D_{\rm eff}(k^2)}{(-(q-k)^2)^{1-D/2}}.
\end{equation}
It is therefore obvious how to calculate the contribution $V_{b11}(q^2)$
from the self energy correction of the massive line and its spectral density
$\rho_{b11}(s)=\slq\rho_{b11}^q(s)+m\rho_{b11}^m(s)$. After integrating by 
parts the final result can be seen to be a convolution of the leading order
contribution with specified weight functions
\begin{eqnarray}\label{massint}
\rho_{b11}^q(s)&=&\frac{g_s^2G(1,1)(D-2)s^{D-2}}{16(4\pi)^{3D/2}(D-1)^2}
  \int_z^1\left(-\frac{\hat\rho_a(x)}{x(1-x)}+2\hat L_b(x)\frac{d}{dx}\right)
  \hat\rho_{a1}(z/x)dx,\nonumber\\
\rho_{b11}^m(s)&=&\frac{g_s^2G(1,1)(D-2)s^{D-2}}{16(4\pi)^{3D/2}(D-1)^2}
  \int_z^1\left(-\frac{\hat\rho_{a+b}(x)}{x(1-x)}+2\hat L_b(x)\frac{d}{dx}
  \right)\hat\rho_{a1}(z/x)dx\qquad\quad
\end{eqnarray}
where
\begin{equation}
\rho_a(s)=:\frac{g_s^2}{(4\pi)^{D/2}}\hat\rho_a(m^2/s),\qquad
\rho_{a+b}(s)=:\frac{g_s^2}{(4\pi)^{D/2}}\hat\rho_{a+b}(m^2/s)
\end{equation}
and
\begin{equation}
L_b(s)=\int_s^\infty\frac{m^2\rho_b(s_1)}{(s_1-m^2)^2}ds_1
  =:\frac{g_s^2}{(4\pi)^{D/2}}\hat L_b(m^2/s).
\end{equation}
The corresponding spectral functions are
\begin{eqnarray}
\hat\rho_{b11}^1(z)&=&\frac{49}{36}+\frac{116}{27}z-\frac{74}9z^2
  +\frac{28}9z^3-\frac{59}{108}z^4
  +\left(\frac13+\frac{44}9z-4z^2+\frac43z^3-\frac29z^4\right)\ln z
  +\strut\nonumber\\&&\strut
  -\left(\frac23+\frac{20}9z-4z^2+\frac43z^3-\frac29z^4\right)\ln(1-z)
  +2z\left(\Li_2(z)-\Li_2(1)+\frac12\ln^2z\right),\nonumber\\[12pt]
\hat\rho_{b11}^2(z)&=&\frac{55}{144}+\frac{329}{54}z-\frac{46}9z^2
  -\frac32z^3+\frac{59}{432}z^4
  +\left(\frac1{12}+\frac{34}9z+\frac32z^2-\frac23z^3+\frac1{18}z^4\right)
  \ln z+\strut\nonumber\\&&\strut
  -\left(\frac16+\frac{22}9z-2z^2-\frac23z^3+\frac1{18}z^4\right)\ln(1-z)
  +\strut\nonumber\\&&\strut
  +\frac23z(2+3z)\left(\Li_2(z)-\Li_2(1)+\frac12\ln^2z\right),\nonumber\\[12pt]
\hat\rho_{b11}^3(z)&=&-\frac{77}{1800}-\frac{433}{360}z+\frac{289}{90}z^2
  -\frac{299}{90}z^3+\frac{613}{360}z^4-\frac{623}{1800}z^5
  +\strut\nonumber\\&&\strut
  -\left(\frac1{30}+\frac56z-\frac43z^2+\frac43z^3-\frac23z^4
  +\frac2{15}z^5\right)\ln z+\strut\nonumber\\&&\strut
  -\left(\frac2{15}-\frac23z+\frac43z^2-\frac43z^3+\frac23z^4
  -\frac2{25}z^5\right)\ln(1-z),\nonumber\\[12pt]
\hat\rho_{b11}^4(z)&=&\hat\rho_{b11}^5(z)\ =\ \hat\rho_{b11}^6(z)
  \ =\ -\frac{107}{7200}-\frac{709}{720}z-\frac{173}{1080}z^2
  +\frac{91}{60}z^3-\frac{199}{480}z^4+\frac{623}{10800}z^5
  +\kern-7pt\strut\nonumber\\&&\strut
  -\left(\frac1{120}+\frac5{12}z+\frac{47}{36}z^2-\frac23z^3+\frac{z^4}6
  -\frac{z^5}{45}\right)\ln z+\strut\nonumber\\&&\strut
  -\left(\frac1{30}-\frac{z}3-\frac29z^2+\frac23z^3-\frac{z^4}6+\frac{z^5}{45}
  \right)\ln(1-z)+\strut\nonumber\\&&\strut
  -\frac23z^2\left(\Li_2(z)-\Li_2(1)+\frac12\ln^2z\right).
\end{eqnarray}
These results have already been renormalized. In addition to
Eq.~(\ref{massint}) we have to take into account a further finite contribution
coming from the singular parts of $a$ and $b$. Since
\begin{equation}
a_s=\frac{-g_s^2}{(4\pi)^{D/2}}\pfrac{\mu^2}{m^2}^\eps\frac G\eps
\Big(1+O(\eps^2)\Big),\qquad
a_s+b_s=\frac{g_s^2}{(4\pi)^{D/2}}\pfrac{\mu^2}{m^2}^\eps\frac G\eps
\Big(2-2\eps+O(\eps^2)\Big)
\end{equation}
we can absorb the singularity into the renormalization factor.
This is the case for the $\msbar$-mass where the finite constants have to be
added to the result. If we absorb the finite constants as well we end
up with the pole mass. This is preferable because then we do not have to take
care of the numerator singularity containing $b$. In any case, we obtain
expressions for $\ln(\mu^2/m^2)$ which have the same coefficients as the poles.

\section{The semi-massive fish contribution (c21)}
\setcounter{equation}{0}\def\theequation{C\arabic{equation}}
In order to determine the semi-massive fish contribution one has to calculate 
a number of scalar two-loop integrals. These so-called {\em prototypes\/}
are spectral functions $\rho_V(n_1,n_2,n_3,n_4,n_5;s/m^2)$. These spectral
functions are determined by the discontinuities of the correlator functions
$V(n_1,n_2,n_3,n_4,n_5;q^2/m^2)$ given by
\begin{eqnarray}
\lefteqn{\frac1{(4\pi)^D}(m^2)^{D-n_1-n_2-n_3-n_4-n_5}
  V(n_1,n_2,n_3,n_4,n_5;q^2/m^2)\ =}\nonumber\\
  &:=&\int\dDk\dDl\frac1{(k^2+m^2)^{n_1}(l^2+m^2)^{n_2}((q-k)^2)^{n_3}
  ((q-l)^2)^{n_4}((k-l)^2)^{n_5}}
\end{eqnarray}
(for convenience written in the Euclidean domain). For later use it is
convenient to use the representation
\begin{equation}
\hat\rho_V(n_1,n_2,n_3,n_4,n_5;z):=z^{D-n_1-n_2-n_3-n_4-n_5}
  \rho_V(n_1,n_2,n_3,n_4,n_5;1/z).
\end{equation}
A subset of the prototypes turn out to be reducible to scalar one-loop
integrals using the spectral representation
$\hat\rho_V(n_1,n_2;z):=z^{D/2-n_1-n_2}\rho_V(n_1,n_2;1/z)$, where the
spectral function $\rho_V(n_1,n_2;s/m^2)$ is given by the discontinuity of the
correlator function $V(n_1,n_2;q^2/m^2)$. One obtains 
\begin{equation}
\frac1{(4\pi)^{D/2}}(m^2)^{D/2-n_1-n_2}V(n_1,n_2;q^2/m^2)
  \ :=\ \int\dDk\frac1{(k^2+m^2)^{n_1}((q-k)^2)^{n_2}}\,.
\end{equation}
An alternative approach is to relate them to the massless one-loop integrals 
$G(n_1,n_2)$ with
\begin{equation}
\frac1{(4\pi)^{D/2}}(q^2)^{D/2-n_1-n_2}G(n_1,n_2)
  \ :=\ \int\dDk\frac1{(k^2)^{n_1}((q-k)^2)^{n_2}}\,.
\end{equation}

\subsection{The proper fish and the spectacle prototype}
\begin{figure}[t]\begin{center}
\epsfig{figure=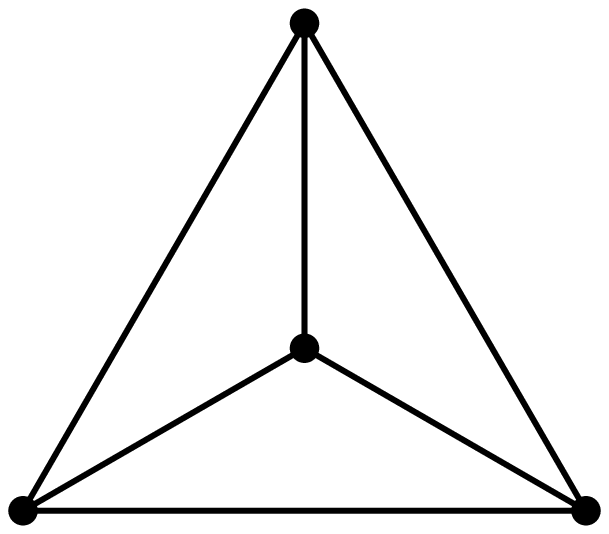,scale=0.5}\quad
\epsfig{figure=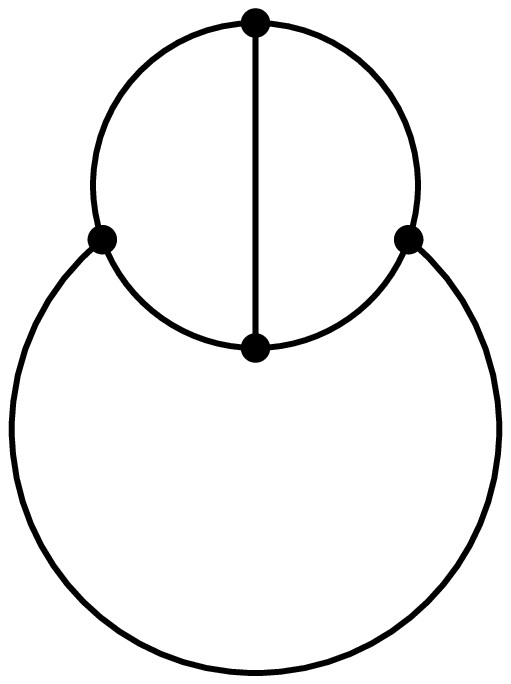,scale=0.5}\qquad
\epsfig{figure=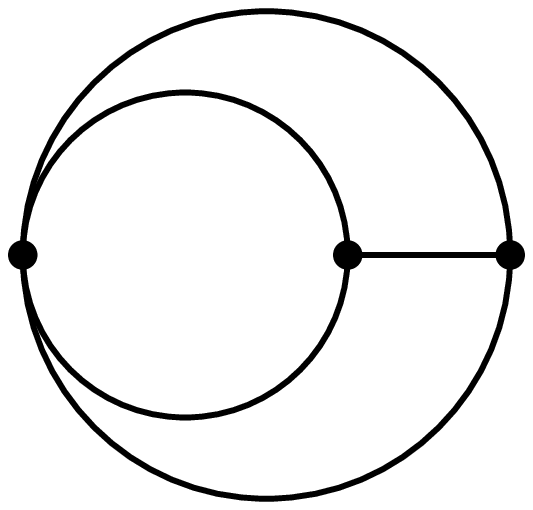,scale=0.5}\quad
\epsfig{figure=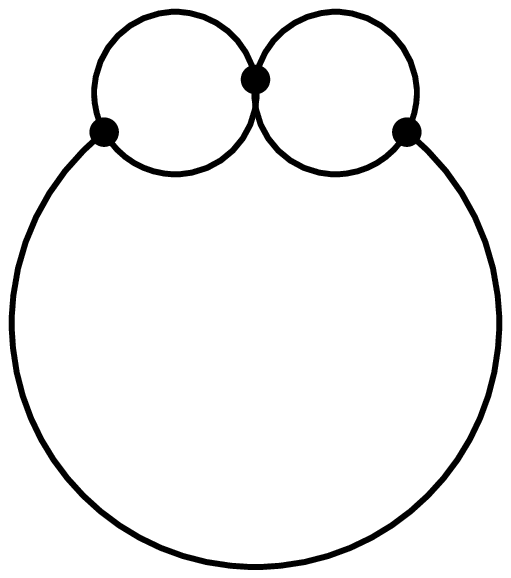,scale=0.5}\\
\strut\kern7pt(a)\kern75pt(b)\kern80pt(c)\kern75pt(d)
\end{center}
\caption{\label{tetra}Different representations of the tetrahedron or
proper fish topology (a-b) and the spectacle topology (c-d)}
\end{figure}

We start with the most difficult prototype $\hat\rho_V(1,1,1,1,1;z)$, the
{\em proper fish} or tetrahedron (see \ Fig.~\ref{tetra}(a) and~(b) for two
different representations of this topology).
The corresponding correlator function can be obtained as a limiting case of an
expression taken from the literature~\cite{Generalis,Broadhurst:1987ei}. The
discontinuity of this expression turns out to be finite. One obtains
\begin{equation}
\hat\rho_V(1,1,1,1,1;z)=4\left(\Li_2(z)+\frac12\ln(1-z)\ln z\right)+O(\eps).
\end{equation}
If the last entry ``1'' is replaced by ``0'', the correlator consists of two
master bubbles. This diagram is termed the {\em spectacle diagram\/}
(cif. Fig.\ref{tetra}(c) and (d)) where the name derives from the pictorial
representation Fig.\ref{tetra}(d). We calculate the spectacle diagram by adding a
further scalar line,
\begin{equation}
\frac{(m^2)^{1-\eps}}{(4\pi)^{D/2}}V(1,1,1,1,0,1;q^2/m^2)
  =\int\dDp\frac1{(q-p)^2}\int V(1,1;p^2/m^2)V(1,1;p^2/m^2).
\end{equation}
Note, however, that we have to subtract the infrared divergence before we can
make use of the dispersive representation. We insert
\begin{equation}
V(1,1;p^2/m^2)=\int\frac{\rho_V(1,1;s/m^2)}{s+p^2}ds
  =V(1,1;-1)-(p^2+m^2)\int\frac{\rho_V(1,1;s/m^2)ds}{(s-m^2)(s+p^2)},
\end{equation}
where we have chosen the momentum subtraction at the point $p^2=-m^2$ in the
Minkowskian domain (cf.\ Appendix~B). One obtains
\begin{eqnarray}
\lefteqn{\frac{(m^2)^{1-\eps}}{(4\pi)^{D/2}}V(1,1,1,1,0,1;q^2/m^2)\ =}\\
  &=&\int\tilde\lambda_a(q^2,s_1)\rho_V(1,1;s_1/m^2)ds_1
  +\int\tilde\lambda_b(q^2,s_1,s_2)\rho_V(1,1;s_1/m^2)\rho_V(1,1;s_2/m^2)
  ds_1ds_2.\nonumber
\end{eqnarray}
The convolution functions are given by
\begin{eqnarray}
\lefteqn{\tilde\lambda_a(q^2,s_1)\ =\ -2\int\dDp
  \frac{V(1,1;-1)(p^2+m^2)}{(s_1-m^2)(s_1+p^2)(q-p)^2}
  \ =\ 2V(1,1;-1)\tilde\lambda_0(q^2,s_1),}\nonumber\\
\lefteqn{\tilde\lambda_b(q^2,s_1,s_2)\ =\ \int\dDp
  \frac{(p^2+m^2)^2}{(s_1-m^2)(s_2-m^2)(s_1+p^2)(s_2+p^2)(q-p)^2}
  \ =}\nonumber\\
  &=&\frac1{(s_1-m^2)(s_2-m^2)(s_2-s_1)}\int\dDp\left(
  \frac{(p^2+m^2)^2}{(s_1+p^2)(q-p)^2}-\frac{(p^2+m^2)^2}{(s_2+p^2)(q-p)^2}
  \right)\ =\nonumber\\
  &=&\frac{(s_1-m^2)^2\tilde\lambda_0(q^2,s_1)
  -(s_2-m^2)^2\tilde\lambda_0(q^2,s_2)}{(s_1-m^2)(s_2-m^2)(s_2-s_1)}
\end{eqnarray}
where
\begin{equation}
\tilde\lambda_0(q^2,s)=\int\dDp\frac1{(s+p^2)(q-p)^2}
  =\frac1{(4\pi)^{D/2}}V(1,1;p^2/s).
\end{equation}
Note that for the reduction to the fundamental convolution function
$\tilde\lambda_0(q^2,s)$ one can use the fact that each non-negative integer 
power of $p^2$ occuring in the integrand of this function can be effectively
replaced by $-s$. Calculating the spectral function we obtain
\begin{eqnarray}
\lefteqn{\frac{(m^2)^{1-\eps}}{(4\pi)^{D/2}}\rho_V(1,1,1,1,0,1;s/m^2)\ =}\\
  &=&\int\lambda_a(s,s_1)\rho_V(1,1;s_1/m^2)ds_1
  +\int\lambda_b(s,s_1,s_2)\rho_V(1,1;s_1/m^2)\rho_V(1,1;s_2/m^2)
  ds_1ds_2\nonumber
\end{eqnarray}
where
\begin{eqnarray}
\lambda_a(s,s_1)&=&2V(1,1;-1)\lambda_0(s,s_1),\qquad
  \lambda_0(s,s_1)\ =\ \frac1{(4\pi)^{D/2}}\rho_V(1,1;s/s_1),\nonumber\\
\lambda_b(s,s_1,s_2)&=&\frac{(s_1-m^2)^2\lambda_0(s,s_1)
  -(s_2-m^2)^2\lambda_0(s,s_2)}{(s_1-m^2)(s_2-m^2)(s_2-s_1)}.
\end{eqnarray}
In order to combine both parts into an integral including a unique convolution
function and a single integrand (which later on will be identified with the
prototype $\rho_V(1,1,1,1,0;s/m^2)$) we use the fact that the second
integral is symmetric in $s_1$ and $s_2$. We further make use of the 
expression
\begin{equation}
\rho_V(1,1;1/z)=\frac{\Gamma(1-\eps)}{\Gamma(2-2\eps)}z^\eps(1-z)^{1-2\eps}
\end{equation}
at $D=4$ (i.e.\ $\eps=0$) for the one-loop spectral function to perform one
of the integrations as a principal value integral. One obtains
\begin{eqnarray}
\lambda_b(s,s_1)&:=&\int\lambda_b(s,s_1,s_2)\rho_V(1,1;s_2/m^2)ds_2
  \ =\nonumber\\
  &=&-2\lambda_0(s,s_1)\frac{s_1-m^2}{s_1}\ln\pfrac{s_1-m^2}{m^2},
\end{eqnarray}
such that
\begin{eqnarray}
\lefteqn{\frac{(m^2)^{1-\eps}}{(4\pi)^{D/2}}\rho_V(1,1,1,1,0,1;s/m^2)
  \ =}\nonumber\\
  &=&\int 2\lambda_0(s,s_1)\left(V(1,1;-1)-\left(1-\frac{m^2}{s_1}\right)
  \ln\left(\frac{s_1}{m^2}-1\right)\right)\rho_V(1,1;s_1/m^2)ds_1.
\end{eqnarray}
We identify
\begin{equation}
\rho_V(1,1,1,1,0;1/z)=2\left(V(1,1;-1)-(1-z)\ln\left(\frac1z-1\right)\right)
  \rho_V(1,1;1/z)
\end{equation}
and finally obtain
\begin{equation}
\hat\rho_V(1,1,1,1,0;z)=2V(1,1;-1)z^{-\eps}\hat\rho_V(1,1;z)
  +2(1-z)^2\left(\ln(1-z)-\ln z\right).
\end{equation}
Next we calculate the prototype $\hat\rho_V(1,1,1,1,-1;z)$. For such
prototypes with negative entries we need the vector integral
$V'(1,1;p^2/m^2)$ defined by
\begin{equation}
\frac1{(4\pi)^{D/2}}(m^2)^{D/2-2}p^\mu V'(1,1;p^2/m^2)
  =\int\dDk\frac{k^\mu}{(k^2+m^2)(p-k)^2}\,.
\end{equation}
One obtains
\begin{equation}
V'(1,1;p^2/m^2)=\frac12\left(1-\frac{m^2}{p^2}\right)V(1,1;p^2/m^2)
  -\frac{m^2}{2p^2}V(1,0;-1).
\end{equation}
Again, we use momentum subtraction at the point $p^2=-m^2$,
\begin{equation}
V'(1,1;p^2/m^2)=V'(1,1;-1)-(p^2+m^2)\int
  \frac{\rho'_V(1,1;s/m^2)ds}{(s+p^2)(s-m^2)}
\end{equation}
where the spectral function is given by
\begin{equation}
\rho'_V(1,1;s/m^2)=\frac12\left(1+\frac{m^2}s\right)\rho_V(1,1;s/m^2),\qquad
\hat\rho'_V(1,1;z)=\frac12(1+z)\hat\rho_V(1,1;z).
\end{equation}
In terms of the above vector integral we calculate
\begin{equation}
V(1,1,1,1,-1;p^2/m^2)=-2V(1,1,1,1,0;p^2/m^2)-2\frac{p^2}{m^2}V'(1,1;p^2/m^2)^2,
\end{equation}
and obtain
\begin{eqnarray}
\hat\rho_V(1,1,1,1,-1;z)&=&-2z\rho_V(1,1,1,1,0;z)+4V'(1,1;-1)z^{-\eps}
  \hat\rho'_V(1,1;z)\,+\nonumber\\[3pt]&&\qquad\qquad\qquad\qquad
  -(1+z)(1-z)^2\left(1+(1+z)\ln\left(\frac1z-1\right)\right)\ =\nonumber\\[3pt]
  &=&-4V(1,1;-1)z^{1-\eps}\hat\rho_V(1,1;z)
  +4V'(1,1;-1)z^{-\eps}\hat\rho'_V(1,1;z)\,+\nonumber\\[7pt]&&\qquad
  -(1+z)(1-z)^2-(1-z)^4\left(\ln(1-z)-\ln z\right).
\end{eqnarray}

\subsection{Prototypes of the class $\hat\rho_V(1,1,0,1,1;z)$}
Prototypes with one vanishing massless propagator reduce to a nested integral.
For the general case that we need to consider here we obtain
\begin{eqnarray}
\lefteqn{\frac1{(4\pi)^D}(m^2)^{D-2-n_4-n_5}V(1,1,0,n_4,n_5)\ =}\nonumber\\
  &=&\int\dDk\frac1{(k^2+m^2)((p-k)^2)^{n_4}}
  \int\dDl\frac1{(l^2+m^2)((k-l)^2)^{n_5}}\ =\nonumber\\
  &=&\frac1{(4\pi)^{D/2}}(m^2)^{D/2-1-n_5}\int\dDk
  \frac{V(1,n_5;k^2/m^2)}{(k^2+m^2)((p-k)^2)^{n_4}}.
\end{eqnarray}
Again making use of momentum subtraction
\begin{equation}
V(1,n_5;k^2/m^2)=V(1,n_5;-1)-(k^2+m^2)\int
  \frac{\rho_V(1,n_5;s/m^2)ds}{(s-m^2)(s+k^2)}
\end{equation}
one ends up with
\begin{eqnarray}
V(1,1,0,n_4,n_5;p^2/m^2)&=&V(1,n_5;-1)V(1,n_4;p^2/m^2)
  -\int\frac{\rho_V(1,n_5;s/m^2)}{s-m^2}V(1,n_4;p^2/s)ds,\nonumber\\
\rho_V(1,1,0,n_4,n_5;s/m^2)&=&V(1,n_5;-1)\rho_V(1,n_4;s/m^2)
  -\int\frac{\rho_V(1,n_5;s_1/m^2)}{s-m^2}V(1,n_4;s/s_1)ds_1\nonumber\\
\end{eqnarray}

\vspace{-24pt}\noindent
and
\begin{eqnarray}
\lefteqn{\hat\rho_V(1,1,0,n_4,n_5;z)\ =}\nonumber\\
  &=&z^{1-n_5-\eps}\left(V(1,n_5;-1)\hat\rho_V(1,n_4;z)
  -\int\frac{\hat\rho_V(1,n_5;z_1)}{z_1^{n_4-n_5+1}(1-z_1)}
  \hat\rho_V(1,n_4;z_1/z)dz_1\right).\qquad\qquad
\end{eqnarray}
Because of $\hat\rho_V(1,0;z)=0$ we obtain
\begin{eqnarray}
\hat\rho_V(1,1,0,1,1;z)&=&V(1,1;-1)z^{-\eps}\hat\rho_V(1,1;z)+1-z+\ln z,
  \nonumber\\[3pt]
\hat\rho_V(1,1,0,1,0;z)&=&V(1,0;-1)z^{1-\eps}\hat\rho_V(1,1;z)+(1-z)z,
  \nonumber\\[3pt]
\hat\rho_V(1,1,0,0,1;z)&=&0
\end{eqnarray}
(furthermore, $\hat\rho_V(1,1,1,0,n_5;z)=\hat\rho_V(1,1,0,1,n_5;z)$ because of
the symmetry of the problem). For the last prototype of this class, 
$\hat\rho_V(1,1,-1,1,1;z)$,
the vector integral $V'$ will appear again. Using a dispersion
relation we obtain
\begin{eqnarray}
\lefteqn{\hat\rho_V(1,1,-1,1,1;z)+(1+z)\rho_V(1,1,0,1,1;z)\ =}\nonumber\\
  &=&2z^{-\eps}V'(1,1;-1)\hat\rho'_V(1,1;z)
  -2\int_z^1\frac{\hat\rho'_V(1,1;z_1)}{z_1(1-z_1)}\hat\rho'_V(1,1;z/z_1)dz_1.
\end{eqnarray}
The final result reads
\begin{eqnarray}
\hat\rho_V(1,1,-1,1,1;z)&=&(1+z)V(1,1;-1)z^{-\eps}\hat\rho_V(1,1;z)
  +2V'(1,1;-1)z^{-\eps}\hat\rho_V(1,1;z)\,+\nonumber\\&&\qquad\qquad\qquad
  -\frac54+z+\frac14z^2-\left(\frac12+z\right)\ln z.
\end{eqnarray}

\subsection{Prototypes of the class $\hat\rho_V(0,1,1,1,1;z)$}
If one of the massive propagators vanishes, the result is given by the product
of a massive and a massless one-loop correlator. The spectral function reads
\begin{equation}
\hat\rho_V(0,n_2,n_3,n_4,n_5;z)=G(n_3,n_5)\hat\rho_V(n_2,n_3+n_4+n_5-D/2;z).
\end{equation}
For the special cases that occur in our calculations we obtain
\begin{eqnarray}
\hat\rho_V(0,1,1,1,1;z)&=&G(1,1)\hat\rho_V(1,\eps+1;z),\nonumber\\[7pt]
\hat\rho_V(0,1,1,0,1;z)&=&G(1,1)\hat\rho_V(1,\eps;z),\nonumber\\[7pt]
\hat\rho_V(0,1,1,-1,1;z)&=&G(1,1)\hat\rho_V(1,\eps-1;z),
\end{eqnarray}
as well as $\hat\rho_V(0,0,n_3,n_4,n_5;z)=0$. The last prototype
$\hat\rho_V(-1,1,1,0,1;z)$ is more difficult. With a little bit of work
one finds 
\begin{equation}
\hat\rho_V(-1,1,1,0,1;z)=-\frac12G(1,1)\left((1-z)\hat\rho_V(1,\eps;z)
  +\hat\rho_V(1,\eps-1;z)\right).
\end{equation}
In order to calculate the final result, we consider the remaining spectral
functions
\begin{eqnarray}
\hat\rho_V(1,\eps-1;z)&=&\frac1{\Gamma(\eps-1)\Gamma(3-2\eps)}\int_z^1
  (1-x)^{2-2\eps}x^{\eps-2}(x-z)^{2-2\eps}dx,\nonumber\\
\hat\rho_V(1,\eps;z)&=&\frac1{\Gamma(\eps)\Gamma(2-2\eps)}\int_z^1
  (1-x)^{1-2\eps}x^{\eps-1}(x-z)^{1-2\eps}dx,\nonumber\\
\hat\rho_V(1,\eps+1;z)&=&\frac1{\Gamma(1+\eps)\Gamma(1-2\eps)}\int_z^1
  (1-x)^{-2\eps}x^\eps(x-z)^{-2\eps}dx.
\end{eqnarray}
The first two integrals can be evaluated for $\eps=0$, while for the last
member of this family the singularity in $G(1,1)$ is not cancelled. However,
we can subtract and add $\hat\rho_V(1,1;z)$ to separate the singular
and finite parts. Using
\begin{equation}
\hat\rho_V(1,1;z)=\frac1{\Gamma(1-\eps)}\int_z^1(1-x)^{-\eps}(x-z)^{-\eps}dx
\end{equation}
and $\Gamma(1+\eps)\Gamma(1-2\eps)=\Gamma(1-\eps)+O(\eps^2)$, we obtain
\begin{eqnarray}
\lefteqn{\frac1\eps\left(\hat\rho_V(1,\eps+1;z)-z^{-\eps}\hat\rho_V(1,1;z)
  \right)\ =}\nonumber\\
  &=&\frac1\eps\int_z^1(1-x)^{-\eps}(x-z)^{-\eps}\left((1-x)^{-\eps}x^\eps
  (x-z)^{-\eps}-z^{-\eps}\right)dx+O(\eps)\ =\nonumber\\
  &=&\int_z^1\left(\ln z-\ln(1-x)+\ln x-\ln(x-z)\right)dx+O(\eps)
  \ =\nonumber\\[7pt]
  &=&1-z+(1-2z)\ln z-2(1-z)\ln(1-z)+O(\eps).
\end{eqnarray}

\subsection{Table containing all needed prototypes}
All necessary prototypes are listed in this subsection, starting from the most
complicated one, the proper fish prototype, to those that are zero. Using
$G(1,1)=G/\eps$ and
\begin{eqnarray}
V(1,1;-1)&=&\frac{\Gamma(\eps)}{1-2\eps}\ =\ \frac G\eps+O(\eps),\nonumber\\
V(1,0;-1)&=&\Gamma(\eps-1)\ =\ -\frac G\eps+1+O(\eps),\nonumber\\
V'(1,1;-1)&=&V(1,1;-1)+\frac12V(1,0;-1)\ =\ \frac G{2\eps}+\frac12+O(\eps),
\end{eqnarray}
one has (in addition to inherent symmetries)
\begin{eqnarray}
\hat\rho_V(1,1,1,1,1;z)&=&4\left(\Li_2(z)+\frac12\ln(1-z)\ln z\right),
  \nonumber\\
\hat\rho_V(1,1,1,1,0;z)&=&2\frac G\eps z^{-\eps}\hat\rho_V(1,1;z)
  -2(1-z)^2\left(\ln(1-z)-\ln z\right),\nonumber\\
\hat\rho_V(1,1,1,1,-1;z)&=&(1-3z)\frac G\eps z^{-\eps}\hat\rho_V(1,1;z)
  +(1-z^2)z-(1-z)^4\left(\ln(1-z)-\ln z\right),\!\nonumber\\
\hat\rho_V(1,1,1,0,0;z)&=&-\frac G\eps z^{1-\eps}\rho_V(1,1;z)+(1-z)z,
  \nonumber\\
\hat\rho_V(1,1,-1,1,1;z)&=&-\frac12(1+z)\frac G\eps z^{-\eps}\hat\rho_V(1,1;z)
  -\frac34+z-\frac14z^2-\left(\frac12+z\right)\ln z,\nonumber\\
\hat\rho_V(1,1,0,1,1;z)
  &=&\frac G\eps z^{-\eps}\hat\rho_V(1,1;z)+1-z+\ln z,\nonumber\\
\hat\rho_V(0,1,1,1,1;z)&=&\frac G\eps z^{-\eps}\hat\rho_V(1,1;z)
  +1-z+(1-2z)\ln z-2(1-z)\ln(1-z),\nonumber\\
\hat\rho_V(0,1,1,0,1;z)&=&\frac{1-z^2}2+z\ln z,\nonumber\\[3pt]
\hat\rho_V(0,1,1,-1,1;z)
  &=&-\frac12\left(\frac13+3z-3z^2-\frac13z^3+2z(1+z)\ln z\right),\nonumber\\
\hat\rho_V(-1,1,1,0,1;z)&=&-\frac16+z-\frac12z^2-\frac13z^3+z^2\ln z
\end{eqnarray}
while
\begin{eqnarray}
\hat\rho_V(1,1,0,0,1;z)&=&\hat\rho_V(0,1,1,1,0;z)\ =\ \hat\rho_V(0,1,0,1,1;z)
  \ =\ \hat\rho_V(0,0,1,1,1;z)\ =\ 0,\nonumber\\[7pt]
\hat\rho_V(1,1,0,0,0;z)&=&\hat\rho_V(0,1,1,0,0;z)\ =\ \hat\rho_V(0,0,1,1,0;z)
  \ =\ \hat\rho_V(0,0,0,1,1;z)\ =\ 0,\nonumber\\[7pt]
\hat\rho_V(0,1,0,0,1;z)&=&\hat\rho_V(0,1,0,1,0;z)\ =\ 0.
\end{eqnarray}
The symmetries are given by
\begin{equation}
\hat\rho_V(n_2,n_1,n_4,n_3,n_5;z)=\hat\rho_V(n_1,n_2,n_3,n_4,n_5;z).
\end{equation}

\subsection{Spectral functions for the semi-massive fish}
The spectral functions that we have obtained in the course of our calculation
are
\begin{eqnarray}
\lefteqn{\hat\rho_{c21}^1(z)\ =\ \frac{49}{36}+\frac{67}{54}z
  -\frac{569}{108}z^2+\frac{169}{54}z^3-\frac{25}{54}z^4
  +\left(\frac13+\frac{14}9z-\frac{43}9z^2+\frac{32}9z^3
  -\frac{25}{54}z^4\right)\ln z+\strut}\nonumber\\&&\strut
  -\left(\frac{31}{18}+\frac{26}{27}z-6z^2+\frac{34}9z^3
  -\frac{25}{54}z^4\right)\ln(1-z)+\strut\nonumber\\&&\strut
  +\left(\frac43+\frac{40}9z-8z^2+\frac83z^3-\frac49z^4\right)
  \left(\Li_2(z)+\frac12\ln z\ln(1-z)\right)+\strut\nonumber\\&&\strut
  +\frac43\left(2z-z^3+\frac{z^4}6\right)\left(\Li_2(z)-\Li_2(1)
  +\frac12\ln^2z\right)
  +8z\left(\Li_3(z)-\Li_3(1)-\frac13\ln z\Li_2(z)\right),\nonumber\\[12pt]
\lefteqn{\hat\rho_{c21}^{1\prime}(z)\ =\ -\frac{25}{144}-\frac{13}{54}z
  +\frac{49}{36}z^2-\frac{17}{18}z^3-\frac{z^4}{432}
  -\left(\frac1{12}+\frac59z-2z^2+\frac89z^3+\frac{z^4}9\right)\ln z
  +\strut}\nonumber\\&&\strut
  +\left(\frac19+\frac89z-2z^2+\frac89z^3+\frac{z^4}9\right)\ln(1-z)
  -\frac23z(1-z^2)\left(\Li_2(z)-\Li_2(1)+\frac12\ln^2z\right),
  \nonumber\\[12pt]
\lefteqn{\hat\rho_{c21}^2(z)\ =\ \frac{55}{144}+\frac{665}{216}z
  -\frac{895}{432}z^2-\frac{325}{216}z^3+\frac{25}{216}z^4
  +\strut}\nonumber\\&&\strut
  +\left(\frac1{12}+\frac{35}{18}z-\frac{77}{36}z^2-\frac{11}6z^3
  +\frac{25}{216}z^4\right)\ln z+\strut\nonumber\\&&\strut
  -\left(\frac{37}{72}+\frac{115}{27}z-3z^2-\frac{17}9z^3
  +\frac{25}{216}z^4\right)\ln(1-z)+\strut\nonumber\\&&\strut
  +\left(\frac13+\frac{44}9z-4z^2-\frac43z^3+\frac{z^4}9\right)
  \left(\Li_2(z)+\frac12\ln z\ln(1-z)\right)+\strut\nonumber\\&&\strut
  +\left(\frac43z+5z^2+\frac23z^3-\frac{z^4}{18}\right)\left(\Li_2(z)
  -\Li_2(-1)+\frac12\ln^2z\right)+\strut\nonumber\\&&\strut
  +(4z+6z^2)\left(\Li_3(z)-\Li_3(1)-\frac13\ln z\Li_2(z)\right),
  \nonumber\\[12pt]
\lefteqn{\hat\rho_{c21}^{2\prime}(z)\ =\ -\frac{31}{576}-\frac{41}{54}z
  +\frac5{16}z^2+\frac{z^3}2+\frac{z^4}{1728}-\left(\frac1{48}+\frac{11}{18}z
  -\frac{z^2}8-\frac79z^3-\frac{z^4}{36}\right)\ln z+\strut}\nonumber\\&&\strut
  +\left(\frac1{36}+\frac79z-\frac79z^3-\frac{z^4}{36}\right)\ln(1-z)
  -\frac z3\left(1+3z+z^2\right)\left(\Li_2(z)-\Li_2(1)+\frac12\ln^2z\right),
  \nonumber\\[12pt]
\lefteqn{\hat\rho_{c21}^3(z)\ =\ \frac{257}{900}-\frac{349}{900}z
  +\frac{1273}{900}z^2-\frac{1321}{300}z^3+\frac{1507}{450}z^4
  -\frac{58}{225}z^5+\strut}\nonumber\\&&\strut
  +\left(\frac1{15}-\frac{z}5+\frac{41}{15}z^2-\frac{27}5z^3+\frac{101}{30}z^4
  -\frac{17}{450}z^5\right)\ln z+\strut\nonumber\\&&\strut
  -\left(\frac{13}{50}-\frac32z+\frac{34}9z^2-6z^3+\frac72z^4-\frac{17}{450}z^5
  \right)\ln(1-z)+\strut\nonumber\\&&\strut
  +\left(\frac4{15}-\frac43z+\frac83z^2-\frac83z^3+\frac43z^4-\frac4{15}z^5
  \right)\left(\Li_2(z)+\frac12\ln z\ln(1-z)\right)+\strut\nonumber\\&&\strut
  +\left(-2z^4+\frac2{15}z^5\right)\left(\Li_2(z)-\Li_2(1)+\frac12\ln^2z
  \right),\nonumber\\[12pt]
\lefteqn{\hat\rho_{c21}^{3\prime}(z)\ =\ -\frac{107}{3600}+\frac{77}{720}z
  -\frac{107}{360}z^2+\frac7{10}z^3-\frac{347}{720}z^4+\frac7{3600}z^5
  +\strut}\nonumber\\&&\strut
  -\left(\frac1{60}-\frac{z}{12}+\frac{z^2}3-\frac23z^3+\frac{13}{36}z^4
  +\frac{z^4}{15}\right)\ln z+\strut\nonumber\\&&\strut
  +\left(\frac1{60}-\frac{z}9+\frac{z^2}3-\frac23z^3+\frac{13}{36}z^4
  +\frac{z^4}{15}\right)\ln(1-z)
  +\frac13z^4\left(\Li_2(z)-\Li_2(1)+\frac12\ln^2z\right),\nonumber\\[12pt]
\lefteqn{\hat\rho_{c21}^4(z)\ =\ \hat\rho_{c21}^5(z)\ =\ \frac{287}{3600}
  -\frac{124}{225}z-\frac{16871}{10800}z^2+\frac{3097}{1350}z^3
  -\frac{1639}{5400}z^4+\frac{29}{675}z^5+\strut}\nonumber\\&&\strut
  +\left(\frac1{60}-\frac2{15}z-\frac{217}{180}z^2+\frac{88}{45}z^3
  +\frac{z^4}{120}+\frac{17}{2700}z^5\right)\ln z+\strut\nonumber\\&&\strut
  -\left(\frac{97}{1800}-\frac{31}{36}z-\frac{35}{27}z^2+\frac{19}9z^3
  -\frac{z^4}{72}+\frac{17}{2700}z^5\right)\ln(1-z)+\strut\nonumber\\&&\strut
  +\left(\frac1{15}-\frac23z-\frac49z^2+\frac43z^3-\frac{z^4}3+\frac2{45}z^5
  \right)\left(\Li_2(z)+\frac12\ln z\ln(1-z)\right)+\strut\nonumber\\&&\strut
  +\left(-\frac53z^2-\frac43z^3+\frac{z^4}6-\frac{z^5}{45}\right)
  \left(\Li_2(z)-\Li_2(1)+\frac12\ln^2z\right)+\strut\nonumber\\&&\strut
  -2z^2\left(\Li_3(z)-\Li_3(1)-\frac13\ln z\Li_2(z)\right),\nonumber\\[12pt]
\lefteqn{\hat\rho_{c21}^{4\prime}(z)\ =\ \hat\rho_{c21}^{5\prime}(z)
  \ =\ \hat\rho_{c21}^{6\prime}(z)\ =\ -\frac{137}{14400}+\frac{131}{1440}z
  -\frac{z^2}{4320}-\frac{149}{720}z^3+\frac{121}{960}z^4-\frac7{21600}z^5
  +\strut}\nonumber\\&&\strut
  -\left(\frac1{240}-\frac{z}{24}-\frac7{72}z^2+\frac{z^3}3-\frac{19}{144}z^4
  -\frac{z^5}{90}\right)\ln z+\strut\nonumber\\&&\strut
  +\left(\frac1{240}-\frac{z}{18}-\frac5{36}z^2+\frac{z^3}3-\frac{19}{144}z^4
  -\frac{z^5}{90}\right)\ln(1-z)+\strut\nonumber\\&&\strut
  +\frac{z^2}{12}(2-z^2)\left(\Li_2(z)-\Li_2(1)+\frac12\ln^2z\right),
  \nonumber\\[12pt]
\lefteqn{\hat\rho_{c21}^6(z)\ =\ \frac{287}{3600}-\frac{221}{900}z
  -\frac{3821}{10800}z^2+\frac{3569}{2700}z^3-\frac{1141}{1350}z^4
  +\frac{29}{675}z^5+\strut}\nonumber\\&&\strut
  +\left(\frac1{60}-\frac2{15}z+\frac{53}{180}z^2+\frac{98}{45}z^3
  -\frac{367}{360}z^4+\frac{17}{2700}z^5\right)\ln z+\strut\nonumber\\&&\strut
  -\left(\frac{49}{600}-\frac{13}{12}z-\frac8{27}z^2+\frac73z^3
  -\frac{25}{24}z^4+\frac{17}{2700}z^5\right)\ln(1-z)+\strut\nonumber\\&&\strut
  +\left(\frac1{15}-\frac23z-\frac49z^2+\frac43z^3-\frac{z^4}3+\frac2{45}z^5
  \right)\left(\Li_2(z)+\frac12\ln z\ln(1-z)\right)+\strut\nonumber\\&&\strut
  +\left(-\frac53z^2+\frac{z^4}2-\frac{z^5}{45}\right)\left(\Li_2(z)-\Li_2(1)
  +\frac12\ln^2z\right)+\strut\nonumber\\&&\strut
  -2z^2\left(\Li_3(z)-\Li_3(1)-\frac12\ln z\Li_2(z)\right).
\end{eqnarray}

\end{appendix}

\end{document}